\documentclass[openacc]{rstransa}%%%%where rstrans is the template name
\usepackage[sort&compress,numbers]{natbib}
\usepackage{hyperref}%\usepackage{hyperref}
\usepackage{amsmath,amssymb,bm} 
\usepackage{graphicx}  
\newcommand{\beq}{\begin{equation}}
\newcommand{\eeq}{\end{equation}}

\def\bea{\begin{eqnarray}}
\def\eea{\end{eqnarray}}

\begin{document}

%%%% Article title to be placed here
\title{Emergent gauge fields
and the
high temperature superconductors}

\author{%%%% Author details
Subir Sachdev}

%%%%%%%%% Insert author address here
\address{Department of Physics, Harvard University, Cambridge, Massachusetts, 02138, USA.\\
Perimeter Institute for Theoretical Physics,\\ Waterloo, Ontario N2L 2Y5, Canada}

%%%% Subject entries to be placed here %%%%
\subject{Condensed matter physics, superconductivity, magnetism}

%%%% Keyword entries to be placed here %%%%

%%%% Insert corresponding author and its email address}
\corres{
\email{sachdev@g.harvard.edu}}

%%%% Abstract text to be placed here %%%%%%%%%%%%
\begin{abstract}
The quantum entanglement of many states of matter can be represented by electric and magnetic fields, much 
like those found in Maxwell's theory. These fields `emerge' from the quantum
structure of the many-electron state, rather than being fundamental degrees of freedom
of the vacuum. I review basic aspects of the theory of emergent gauge fields in insulators
in an intuitive manner. In metals, Fermi liquid theory relies on adiabatic 
continuity from the free electron state, and its central consequence is the existence of long-lived
electron-like quasiparticles around a Fermi surface enclosing a volume determined by the total density of electrons,
via the Luttinger theorem. However long-range entanglement and emergent gauge fields
can also be present in metals. I focus on the `fractionalized Fermi liquid' (FL*) state, which also has
long-lived electron-like quasiparticles around a Fermi surface; however the Luttinger theorem on 
the Fermi volume is violated, and this requires the presence of emergent gauge fields, and the associated
loss of adiabatic continuity to the free electron state. Finally, I present a brief survey of some
recent experiments in the hole-doped cuprate superconductors, and interpret the properties
of the pseudogap regime in the framework of the FL* theory.
\end{abstract}
%%%%%%%%%%%%%%%%%%%%%%%%%%%
%%%%%%%%%%%%%%% End of first page %%%%%%%%%%%%%%%%%%%%%
\begin{fmtext}
~\\
~\\
Talk presented at {\em Unifying physics and technology in light of Maxwell's equations\/}, Discussion Meeting at the Royal Society, London, November 16-17, 2015, celebrating the 150th anniversary of Maxwell's equations.\\~\\
\href{http://arxiv.org/abs/1512.00465}{\large arXiv:1512.00465}.
\end{fmtext}

\maketitle

%%%%%%%%%% Insert the texts which can accomdate on firstpage in the tag "fmtext" %%%%%

\section{Introduction}
\label{sec:intro}
%%%% Insert A head here

The copper-based high temperature superconductors have provided a fascinating and fruitful 
environment for the study of quantum correlations in many-electron systems for over two decades. 
Significant experimental and theoretical advances have appeared at a steady pace over the years. In this article, I will
review some theoretical background, and use it to interpret some remarkable recent experiments \cite{Marel13,MG14,Fujita14,Comin14,Forgan15,MHH15a,MHH15b,Badoux15}.
In particular, I argue that modern theoretical ideas on long-range quantum entanglement and emergent gauge
fields provide a valuable framework for understanding the experimental results. I will discuss experimental
signatures of quantum phases with emergent gauge fields, and their connections to the recent observations.

\begin{figure}
\begin{center}
\includegraphics[height=7.5cm]{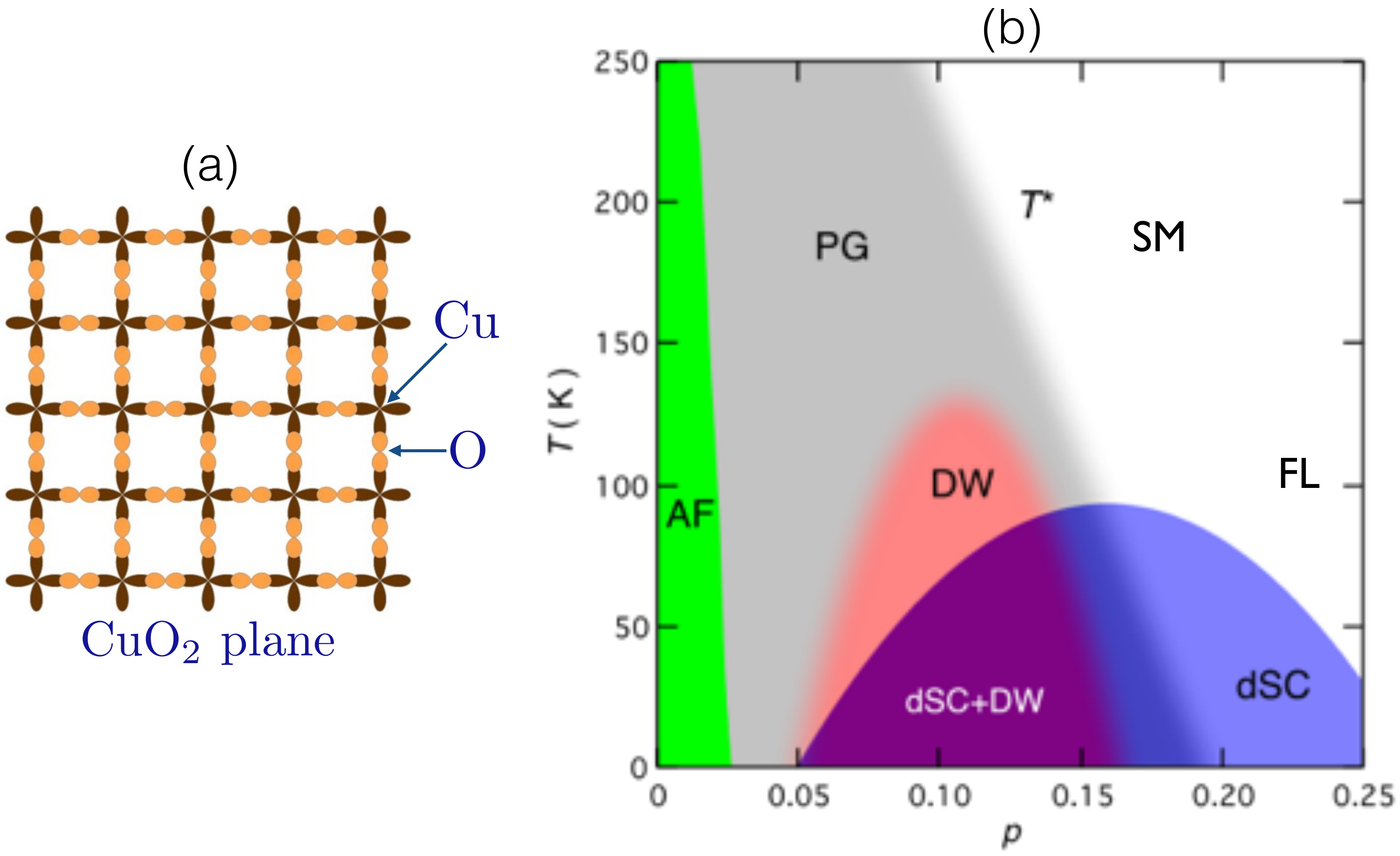}
\end{center}
\caption{(a) The square lattice of Cu and O atoms found in every copper-based high temperature superconductor.
(b) A schematic phase diagram of the YBCO superconductors as a function of the hole density $p$
and the temperature $T$ adapted from Ref.~\cite{MHH15a}. 
The phases are discussed in the text: AF--insulating antiferromagnet, PG--pseudogap, 
DW--density wave, dSC--$d$-wave superconductor, SM--strange metal, FL--Fermi liquid. The critical temperature for superconductivity
is $T_c$, and $T^\ast$ is the boundary of the pseudogap regime.}
\label{fig:phasediag}
\end{figure}

The common feature of all the copper-based superconductors is the presence of a square lattice of Cu and O
atoms shown in Fig.~\ref{fig:phasediag}a. 
For the purposes of this article, we can regard the O $p$ orbitals as filled with pairs
of electrons and inert. Only one of the Cu orbitals is active, and in a parent insulating compound, this orbital has
a density of exactly one electron per site. The rest of this article will consider the physical properties of this Cu orbital
residing on the vertices of a square lattice. It is customary to measure the density of electrons relative to the parent
insulator with one electron per site: we will use $p$ to denote the hole density: {\em i.e.\/} such a state has a density of $1-p$
electrons per Cu site. A recent schematic phase diagram of the hole-doped superconductor YBCO is shown in Fig.~\ref{fig:phasediag}b
as a function of $p$ and the temperature $T$.
The initial interest in these compounds was sparked by the presence of high temperature superconductivity, indicated by the 
large values of $T_c$ in Fig.~\ref{fig:phasediag}b. However, I will not discuss the origin of this superconductivity in this article.
Rather, the focus will be on the other phases, and in particular the pseudogap metal (PG in Fig.~\ref{fig:phasediag}b):
the physical properties of this metal differ qualitatively from those of conventional metals, and so are of significant
intrinsic theoretical interest. Furthermore, superconductivity appears as a low temperature
instability of the pseudogap, so a theory of the high value of $T_c$ can only appear after a theory of the PG metal.

We begin our discussion by describing the simpler phases at the extremes of $p$ in Fig.~\ref{fig:phasediag}b.

\begin{figure}
\begin{center}
\includegraphics[height=5cm]{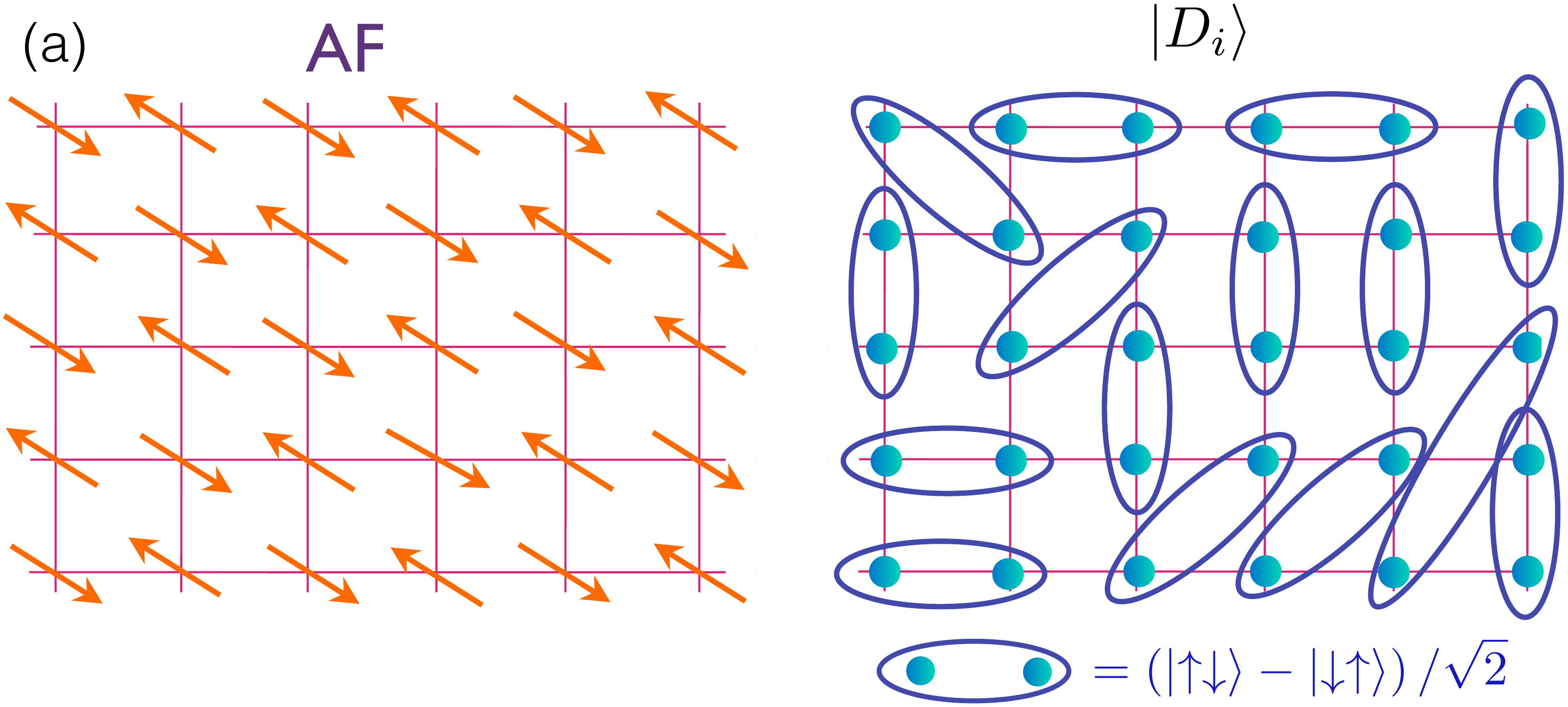}
\end{center}
\caption{(a) The insulating AF state at $p=0$. (b) Component of a `resonating valence bond' wavefunction for the antiferromagnet which preserves
spin rotation symmetry; all the $\left| D_i \right\rangle$ in Eq.~(\ref{eq:rvb}) have similar pairings of electrons on nearby sites (not necessarily nearest-neighbor).}
\label{fig:af}
\end{figure}

At (and near) $p=0$, we have the antiferromagnet (AF) which is sketched in Fig.~\ref{fig:af}a.
The Coulomb repulsion between the electrons keeps their charges immobile on the Cu lattice sites, so that each site has exactly one
electron. The Coulomb interaction is insensitive to the spin of the electron, and so it would appear that each electron spin
is free to rotate independently on each site. However, there are virtual `superexchange' processes which induce terms
in the effective Hamiltonian which prefer opposite orientations of nearest-neighbor spins, and the optimal state turns out
to be the antiferromagnet (AF) sketched in Fig.~\ref{fig:af}a. In this state, the spins are arranged in a checkerboard pattern, so that
all the spins in one sublattice are parallel to each other, and anti-parallel to spins on the other sublattice. Two key features of this
AF state deserve attention here: ({\em i\/}) The state breaks a global spin rotation symmetry, and essentially all of its
low energy properties can be described by well-known quantum field theory methods associated with spontaneously broken
symmetries. ({\em ii\/}) The wavefunction does not have long-range entanglement, and the exact many-electron wavefunction
can be obtained by a series of local unitary transformations on the simple product state sketched in Fig.~\ref{fig:af}a.

At the other end of larger values of $p$, we have the Fermi liquid (FL) phase. This is a metallic state, in which the electronic
properties are most similar to those of simple monoatomic metals like sodium or gold. This is also a quantum state without
long-range entanglement, and the many-electron wavefunction can be well-approximated by a product over single electron momentum eigenstates (Bloch waves); note the contrast from the AF state, where the relevant single-particle states were localized on 
single sites in position space. 
We will discuss some further important properties of the FL state in Section~\ref{sec:fl}.

Section~\ref{sec:sl} will describe possible insulating states on the square lattice, other than the simple AF state found in 
the cuprate compounds at $p=0$. The objective here will be to introduce states with long-range quantum entanglement in 
a simple setting, and highlight their connection to emergent gauge fields. Then Section~\ref{sec:ffl} will combine
the descriptions of Sections~\ref{sec:fl} and~\ref{sec:sl} to propose a metallic state with long-range quantum entanglement
and emergent gauge fields: the fractionalized Fermi liquid (FL*). Finally, in Section~\ref{sec:pg}, we will review
the evidence from recent experiments that the pseudogap (PG) regime of Fig.~\ref{fig:phasediag}b is described by a FL* phase.

I also note here another recent review article \cite{DCSS15}, 
which discusses similar issues at a more specialized level aimed at
condensed matter physicists. The gauge theories of the insulators discussed in Section~\ref{sec:sl} were reviewed in
earlier lectures \cite{SS04,SS10}.

\section{Emergent gauge fields in insulators}
\label{sec:sl}

The spontaneously broken spin rotation symmetry of the AF state at $p=0$ is not observed at higher
$p$. This section will therefore describe quantum states which preserve spin rotation symmetry. However,
in the interests of theoretical simplicity, we will discuss such states in the insulator at the density of $p=0$, and assume
that the AF state can be destabilized by suitable further-neighbor superexchange interactions between the electron spins.

We begin with the `resonating valence bond' (RVB) state
\beq
\left| \Psi \right\rangle = \sum_i c_i \left| D_i \right \rangle \label{eq:rvb}
\eeq
where $i$ extends over all possible pairings of electrons on nearby sites, and a state $\left| D_i \right \rangle$ associated with
one such pairing is shown in Fig.~\ref{fig:af}b; the $c_i$ are complex co-efficients we will leave unspecified here. 
Note that the electrons in a valence bond need not be nearest-neighbors.
Each $\left| D_i \right \rangle$ is a spin singlet, and so spin rotation invariance is preserved; 
the antiferromagnetic exchange interaction is optimized between the electrons within a single valence bond, but not between electrons in separate valence bonds.
We also assume that the $c_i$ respect the translational
and other symmetries of the square lattice. Such a state was first proposed by Pauling \cite{Pauling49} as a description of a simple
metal like lithium. We now know that Pauling's proposal is incorrect for such metals. But we will return to a variant of the
RVB state in Section~\ref{sec:ffl} which does indeed describe a metal, and this metal will be connected to the phase diagram of 
the cuprates in 
Section~\ref{sec:pg}. Anderson revived
the RVB state many years later \cite{Anderson73} as a description of Mott insulators: these are materials with a density of one
electron per site, which are driven to be insulators by the Coulomb repulsion between the electrons (contrary to the Bloch theorem
for free electrons, which requires metallic behavior at this density). 

In a modern theoretical framework, we now realize that the true significance of the Pauling-Anderson RVB proposal was that it was
the first ansatz to realize {\em long-range\/} quantum entanglement.  Similar entanglement appeared 
subsequently in Laughlin's wavefunction for the fractional quantum Hall state \cite{RBL83}, and for RVB states in the absence of time-reversal 
symmetry \cite{RBL87}. The long-range nature of the entanglement can be made precise by
computation of the `topological entanglement entropy' \cite{AKJP05,XGW05,Melko15}. But here we will be satisfied by a 
qualitative description of the sensitivity of the spectrum of states to the topology of the manifold on which the square lattice resides.
The sensitivity is present irrespective of the size of the manifold (provided it is much larger than the lattice spacing), and so indicates
that the information on the quantum entanglement between the electrons is truly long-ranged. A wavefunction which is
a product of localized single-particle states would not care about the global topology of the manifold.

\begin{figure}
\begin{center}
\includegraphics[height=5.5cm]{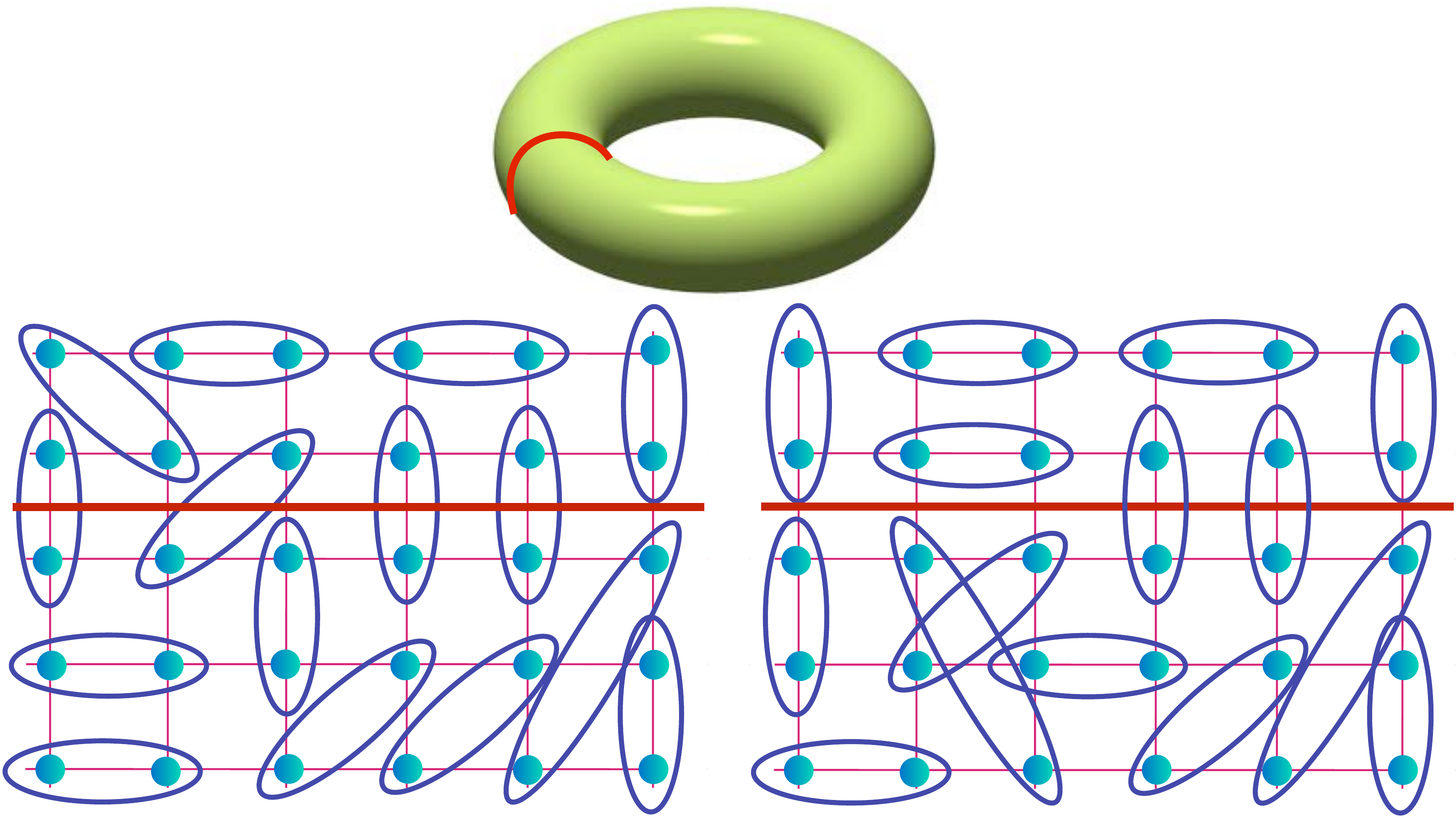}
\end{center}
\caption{Sensitivity of the RVB state to the torus geometry: the number of valence bonds crossing the cut (red line)
can only differ by an even integer between any two configurations (like those shown) which differ by an arbitrary local arrangement of valence bonds.}
\label{fig:torus}
\end{figure}

\begin{figure}
\begin{center}
\includegraphics[height=7.5cm]{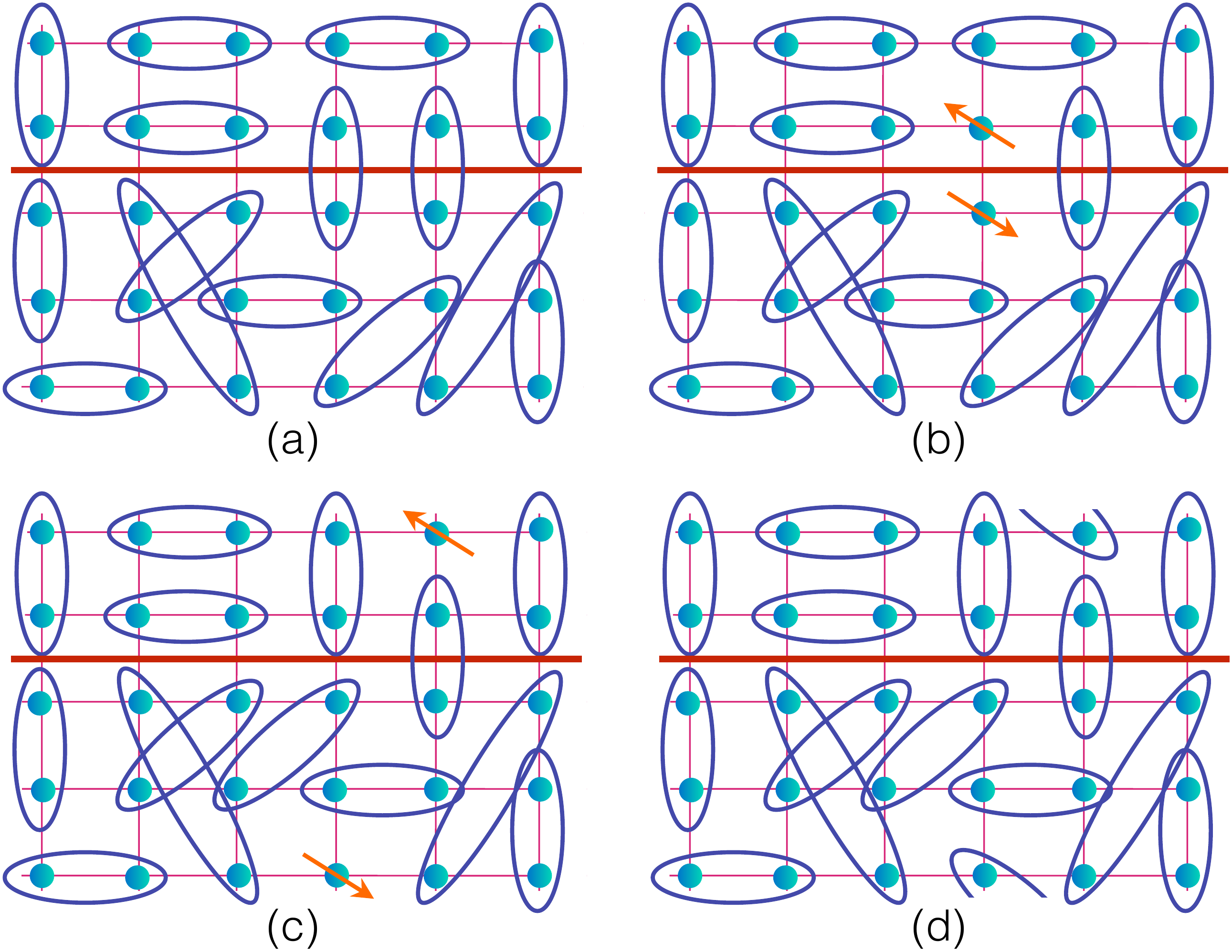}
\end{center}
\caption{Non-local process which changes the parity of the number of valence bonds crossing the cut. A valence bond splits into two spins, which pair up again
after going around the torus.}
\label{fig:cut}
\end{figure}

The basic argument on the long-range quantum information contained in the RVB state is summarized in Fig.~\ref{fig:torus}. 
Place the square lattice on a very large torus ({\em i.e. \/} impose periodic boundary conditions in both directions),
draw an arbitrary imaginary cut across the lattice, indicated by the red line, and count the number of valence bonds crossing the cut.
It is not difficult to see that any {\em local\/} re-arrangement of the valence bonds will preserve the number of valence bonds
crossing the cut modulo 2. Only very non-local processes can change the parity of the valence bonds crossing the cut: one such process
involves breaking a valence bond across the cut into its constituent electrons, and moving the electrons separately around a cycle of the torus crossing the cut, 
so that they meet on the other side and form a new valence bond which no longer crosses the cut--see Fig.~\ref{fig:cut}.
Ignoring this very non-local process, we see that the Hilbert space splits into disjoint sectors, containing states with even or odd number of valence
bonds across the cut \cite{Thouless87,KRS88}. Locally, the two sectors are identical, and so we expect the two sectors to have ground states (and also excited states) of 
nearly the same energy for a large enough torus. The presence of these near-degenerate states is dependent on the global spatial topology, {\em i.e. \/} it requires periodic boundary conditions around the cycles of the torus,  and so can be viewed as a signature
of long-range quantum entanglement.

The above description of topological degeneracy and entanglement relies on a somewhat arbitrary and imprecise trial
wavefunction. A precise understanding is provided by a formulation of the physics of RVB in terms of an emergent gauge theory,
the first example of which was introduced by Baskaran and Anderson \cite{GBPWA88}.
Such a formulation provides another way to view the nearly-degenerate states obtained above on a torus: they
are linear combinations of states obtained by inserting fluxes of the emergent gauge fields through the cycles of the torus. 

\begin{figure}
\begin{center}
\includegraphics[height=4.5cm]{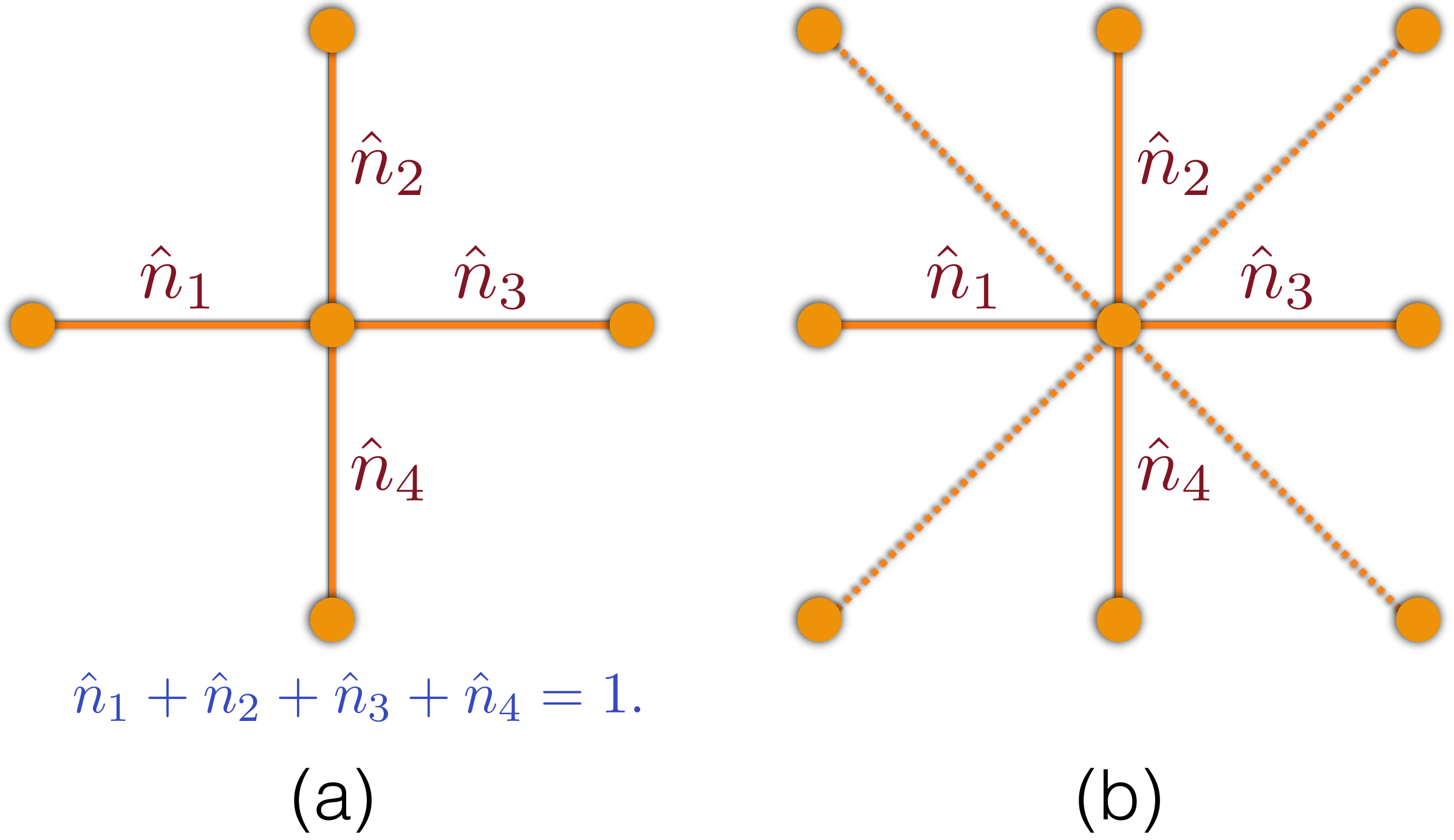}
\end{center}
\caption{(a) Nearest-neighbor valence bond number operators, proportional to the electric field of a compact U(1) gauge theory.
(b) Model with valence bonds connecting the same sublattice: now the constraint on the number operators is modified, and the spin
liquid is described by a $\mathbb{Z}_2$ gauge theory.}
\label{fig:qed}
\end{figure}

\begin{figure}
\begin{center}
\includegraphics[height=6.2cm]{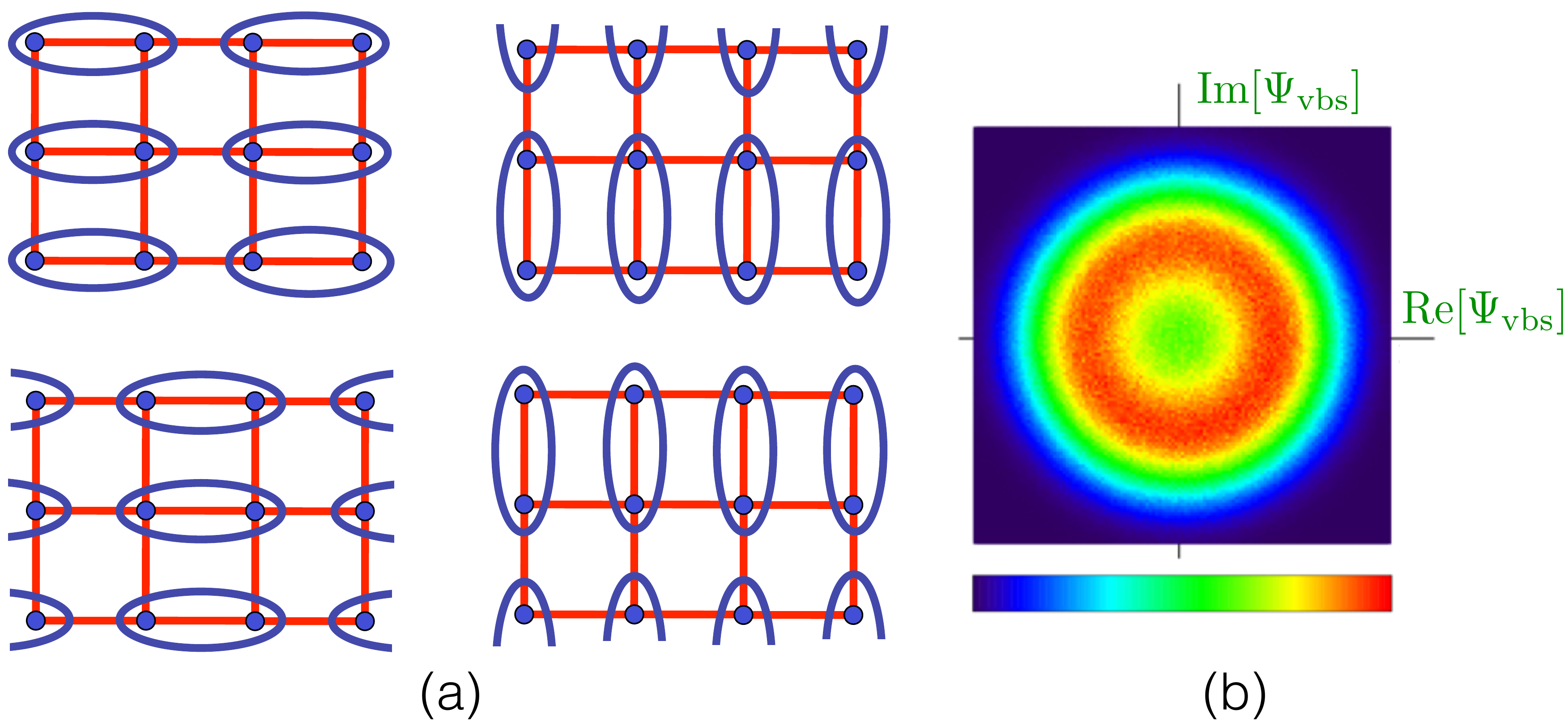}
\end{center}
\caption{(a) The 4 VBS states which break square lattice rotational symmetry
(b) Distribution of the complex VBS order parameter $\Psi_{\rm vbs}$ in the quantum Monte Carlo study by Sandvik \cite{Sandvik07}; the real
and imaginary parts of this order measure the probability of the VBS states in the first and second columns. The near circular distribution of
$\Psi_{\rm vbs}$ reflects an emergent symmetry which is a signature of the existence of a photon.}
\label{fig:vbs}
\end{figure}

The formulation as a gauge theory \cite{GBPWA88,EFSK90} 
becomes evident upon considering a simplified model with valence bonds only
between nearest-neighbor sites on the square lattice. We introduce valence bond number operators $\hat{n}$ on every
nearest-neighbor link, and then there is a crucial constraint that there is exactly one valence bond emerging from
every site, as illustrated in Fig.~\ref{fig:qed}a. 
After introducing oriented `electric field' operators $\hat{E}_{i \alpha} = 
(-1)^{i_x + i_y} \hat{n}_{i \alpha}$ (here $i$ labels sites of the square lattice, and $\alpha = x, y$ labels the two directions), this
local constraint can be written in the very suggestive form \cite{EFSK90}
\beq
\Delta_\alpha \hat{E}_{i \alpha} = \rho_i , \label{eq:gauss}
\eeq
where $\Delta_\alpha$ is a discrete lattice derivative, and $\rho_i \equiv (-1)^{i_x + i_y}$ is a background `charge' density. 
Eq.~(\ref{eq:gauss}) is analogous to Gauss's Law in electrodynamics, and a key indication that the physics of resonating valence
bonds is described by an emergent gauge theory. An important difference from Maxwell's U(1) electrodynamics is that the eigenvalues
of the electric field operator $\hat{E}_{i \alpha}$ must be integers. In terms of the canonically conjugate gauge field $\hat{A}_{i \alpha}$
\beq
[ \hat{A}_{i \alpha}, \hat{E}_{j \beta}] = i \hbar \delta_{ij} \delta_{\alpha\beta},
\eeq
the integral constraint translates into the requirement that $\hat{A}_{i \alpha}$ is a compact angular variable on a unit circle,
and that $\hat{A}_{i \alpha}$ and $\hat{A}_{i \alpha} + 2 \pi$ are equivalent. So there is an equivalence between the quantum 
theory of nearest-neighbor resonating valence bonds on a square lattice, and compact U(1) electrodynamics in the presence of 
fixed background charges $\rho_i$ \cite{EFSK90}. 
A non-perturbative analysis of such a theory shows \cite{NRSS89,NRSS90} that ultimately
there is no gapless `photon' associated with the emergent gauge field $\hat{A}$: compact U(1) electrodynamics is confining
in 2 spatial dimensions, and in the presence of the background charges the confinement leads to valence bond solid (VBS)
order illustrated in Fig.~\ref{fig:vbs}. 
The VBS state breaks square lattice rotation symmetry, and all excitations of the
antiferromagnet, including the incipient photon, have an energy gap. 
In subsequent work, it was realized that the gapless photon can re-emerge at special
`deconfined' critical points \cite{senthil1,AVLBTS04,FHMOS04} or phases \cite{Hermele04}, even in 2 spatial dimensions. In particular, in certain 
models with
a quantum phase transition between a VBS state and the ordered antiferromagnet in Fig.~\ref{fig:af}a \cite{NRSS89,NRSS90,senthil1}, the quantum critical point
supports a gapless photon (along with gapless matter fields). 
This is illustrated in Fig.~\ref{fig:vbs}b by numerical results of Sandvik \cite{Sandvik07}: the circular distribution of valence bonds is evidence for an emergent continuous lattice rotation symmetry, and the associated Goldstone
mode is the dual of the photon.

Although U(1) gauge theory does realize spin liquids with long-range entanglement and emergent photons, the gaplessness 
and `criticality' of the spin liquids indicates the presence of long-range valence bonds, and the Pauling-Anderson trial wavefunctions 
are poor descriptions of such states. 
A stable, gapped quantum state with time-reversal symmetry, long-range entanglement and emergent gauge fields was first established in 
Refs.~\cite{NRSS91,RJSS91,XGW91,SSkagome,MVSS99}
using a model with short-range valence bonds which also connect sites
on the same sublattice (Fig.~\ref{fig:qed}b). It was shown \cite{NRSS91,RJSS91,XGW91,SSkagome,MVSS99} that 
the same-sublattice bonds act like charge $\pm 2$ Higgs fields in the compact U(1) gauge theory, and 
in such gauge theories there can be \cite{EFSS79,Bais92}  a `Higgs' phase.  Such a phase realizes a stable, gapped, RVB state preserving all symmetries of the Hamiltonian, including time-reversal, and is described by an emergent $\mathbb{Z}_2$ gauge theory \cite{RJSS91,MVSS99}. The $\mathbb{Z}_2$ gauge theory can be viewed as a discrete analog of the compact U(1) theory in 
which the gauge field takes only two possible values $\hat{A}_{i \alpha} = 0, \pi$. 
The intimate connection between  a spin liquid with a deconfined $\mathbb{Z}_2$ gauge field, and a non-bipartite RVB trial wavefunction like Eq.~(\ref{eq:rvb}),
was shown convincingly by Wildeboer {\em et al.\/} \cite{Melko15}.
Upon varying parameters in the
underlying Hamiltonian, the $\mathbb{Z}_2$ spin liquid can undergo a confinement transition to a VBS phase which is described
by a dual frustrated Ising model \cite{RJSS91,MVSS99}. Since these early works, the $\mathbb{Z}_2$ spin liquid has appeared
in a number of other models \cite{TSMPAF00,RMSLS01,Kitaev03,Wen03,Freedman04}, including the exactly 
solvable `toric code' \cite{Kitaev03}.

There is an interesting analogy between the theory of $\mathbb{Z}_{2}$ spin liquids 
and that of the $\mathbb{Z}_2$ 
topological insulator proposed more recently by Kane and Mele \cite{KaneMele05}. The $\mathbb{Z}_2$ 
topological insulator is the time-reversal invariant analog of the {\em integer\/} quantum Hall state of fermions at filling
fraction $\nu=1$, and the former is obtained, roughly speaking, by taking two copies of the latter. 
Kalmeyer and Laughlin \cite{RBL87} asserted that RVB states of spin systems must also
break time-reversal symmetry and are analogous to the {\em fractional\/} quantum Hall state of bosons at 
filling fraction $\nu=1/2$. However, it was subsequently shown \cite{NRSS91,RJSS91,XGW91,SSkagome,MVSS99} that a time-reversal invariant RVB state is possible,
and this is the $\mathbb{Z}_2$ spin liquid discussed above. Interestingly, the $\mathbb{Z}_2$ spin 
liquid also has a connection to the theory
of 2 copies of the $\nu=1/2$ quantum Hall state. These connections between quantum Hall states, topological insulators,
and spin liquids can be made precise using abelian Chern-Simons gauge theories \cite{Freedman04,LuVishwanath12}.

\section{The Fermi liquid}
\label{sec:fl}

We now turn to the metallic state found at large $p$ (above the superconducting $T_c$) in Fig.~\ref{fig:phasediag}b. 
This is the familiar Fermi liquid (FL), similar to that found in simple metals like sodium or gold.

\begin{figure}
\begin{center}
\includegraphics[height=4cm]{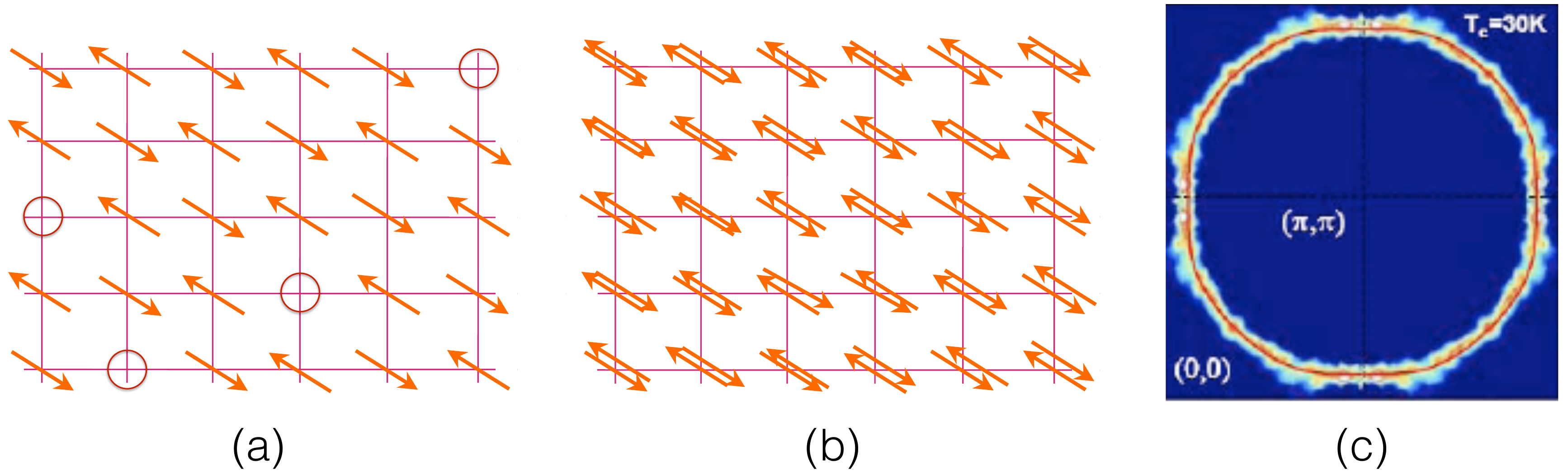}
\end{center}
\caption{(a) State obtained after removing electrons with density $p$ from the AF state in Fig.~\ref{fig:af}a.
Relative to the fully-filled state with 2 electrons per site in (b), this state has a density of holes equal to $1+p$.
(c) Photoemission results from Ref.~\cite{Dama05} showing a Fermi surface of size $1+p$ in the FL region of Fig.~\ref{fig:phasediag}b. This is the expected Luttinger volume at this density, in a state without any antiferromagnetic order.}
\label{fig:fl}
\end{figure}

The key properties of a Fermi liquid, reviewed in many text books are:
\begin{itemize}
\item 
The Fermi liquid state of interacting electrons is adiabatically connected to the free electron state.
The ground state of free electrons has a Fermi surface in momentum space, 
which separates the occupied and empty momentum eigenstates. This Fermi surface is also present
in the interacting electron state, and the low energy excitations are long-lived, electron-like quasiparticles near the 
Fermi surface.
\item The Luttinger theorem states that the volume enclosed by the Fermi surface ({\em i.e.\/} the Fermi volume) is equal (modulo phase space factors we
ignore here) to the total density of electrons. This equality is obviously true for free electrons, but is also proven
to be true, to all orders in the interactions, to any state adiabatically connected to the free electron state.
\item For simple convex Fermi surfaces, the Fermi volume can be measured by the Hall co-efficient, $R_H$, measuring the transverse
voltage across a current in the presence of an applied magnetic field. We have $1/(e R_H) = - (\mbox{electron density})$ for 
electron-like Fermi surfaces, and $1/(e R_H) = (\mbox{hole density})$ for 
hole-like Fermi surfaces.
\end{itemize}
For the cuprates, the Fermi liquid is obtained by removing a density, $p$, of electrons from the insulating antiferromagnet,
as shown in Fig.~\ref{fig:fl}a. 
Relative to the fully-filled state with 2 electrons on each site (Fig.~\ref{fig:fl}b), this state has a density
of holes equal to $1+p$. Hence the Fermi liquid state without antiferromagnetic order 
can have a single hole-like Fermi surface with a Fermi volume of $1+p$
(and {\em not\/} $p$). And indeed, just such a Fermi surface is observed in photoemission experiments in the cuprates
(Fig.~\ref{fig:fl}c)
in the region marked FL in Fig.~\ref{fig:phasediag}b.

\section{Emergent gauge fields in a metal: the fractionalized Fermi liquid}
\label{sec:ffl}

Next, we turn our attention to the smaller $p$ region marked PG (``pseudogap'') in Fig.~\ref{fig:phasediag}b.
We will review the experimental observations in this region in Section~\ref{sec:pg}, but for now we note that in 
many respects this region behaves like an ordinary Fermi liquid, but with the crucial difference that the density
of charge carriers is $p$ and not the Luttinger density of $1+p$. So here we ask the theoretical question: is it possible to obtain
a Fermi liquid which violates the Luttinger theorem and has a Fermi surface of size $p$ of electron-like quasiparticles?
From the structure of the Luttinger theorem we know that any such state cannot be adiabatically connected to the free
electron state. A key result is that long-range quantum entanglement and associated emergent gauge fields are {\em necessary\/}
characteristics of metallic states which violate the Luttinger theorem, and these also break the adiabatic connection
to the free electron state \cite{TSMVSS04,APAV04}.

(There are claims \cite{YRZ} that 
zeros of electron Green's functions can be used to modify the Luttinger result. I believe such results are artifacts of simplified models.
Such zeros do not generically exist as lines in the Brillouin zone (in two dimensions) for gapless states, because both the real
and imaginary parts of the Green's functions have to vanish.) 

\begin{figure}
\begin{center}
\includegraphics[height=4cm]{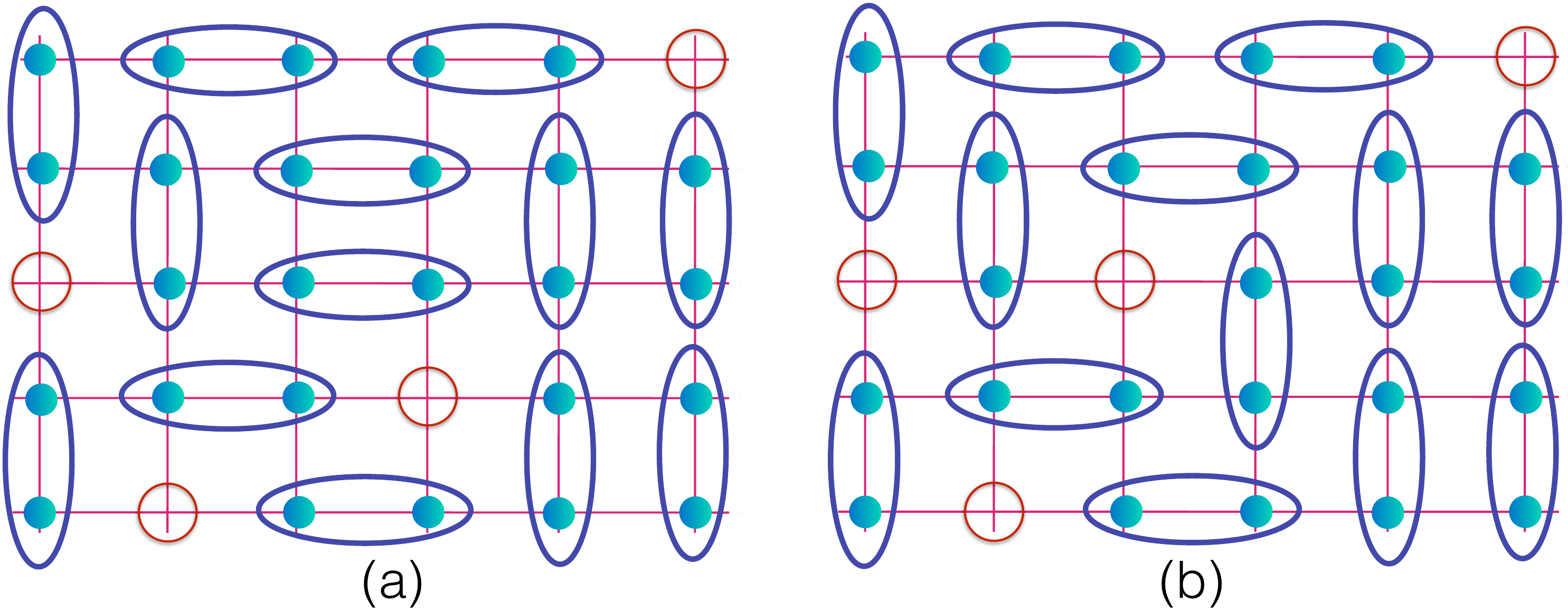}
\end{center}
\caption{(a) State obtained by pairing the spins in Fig.~\ref{fig:fl}a into valence bonds. (b) Resonance between the valence bonds leads to the motion of the vacancy in the center of the figure. The mobile vacancy is a `holon', carrying unit charge but no spin. If the holons have fermionic statistics, such a mobile holon state can realize a holon metal. Only nearest-neighbor pairs of spins are shown for simplicity.}
\label{fig:holon}
\end{figure}

One way to obtain a metal with carrier density $p$, and without antiferromagnetic order is to imagine that the electron spins
in Fig.~\ref{fig:fl}a pair up into resonating valence bonds, rather like the insulator in Fig.~\ref{fig:af}b.
This is illustrated in Fig.~\ref{fig:holon}a; the resonance between the valence bonds can now allow processes in which
the vacant sites can move, as shown in Fig.~\ref{fig:holon}b. 
As this process now transfers physical charge, the resulting
state can be expected to be electrical conductor. A subtle computation is required to determine the quantum statistics obeyed by the mobile vacancies, but depending upon the parameter regimes, it can be either bosonic or fermionic \cite{KRS87,RC89}.
Assuming fermionic statistics, we have the possibility that the vacancies will form a Fermi surface, realizing a metallic state.
Note that the vacancies do not transport spin, and such spinless charge carriers are often referred to as `holons'; the metallic
state we have postulated is a holon metal. The low energy quasiparticles near the Fermi surface of the holon metal will also
be holons, carrying unit electrical charge but no spin. Consequently, such quasiparticles are not directly observable in photoemission
experiments, which necessarily eject bare electrons with both charge and spin. As low energy electronic quasiparticles are observed
in photoemission studies of the PG region in the cuprates (see Section~\ref{sec:pg}), 
the holon metal is not favored as a candidate for the PG metal.

\begin{figure}
\begin{center}
\includegraphics[height=5cm]{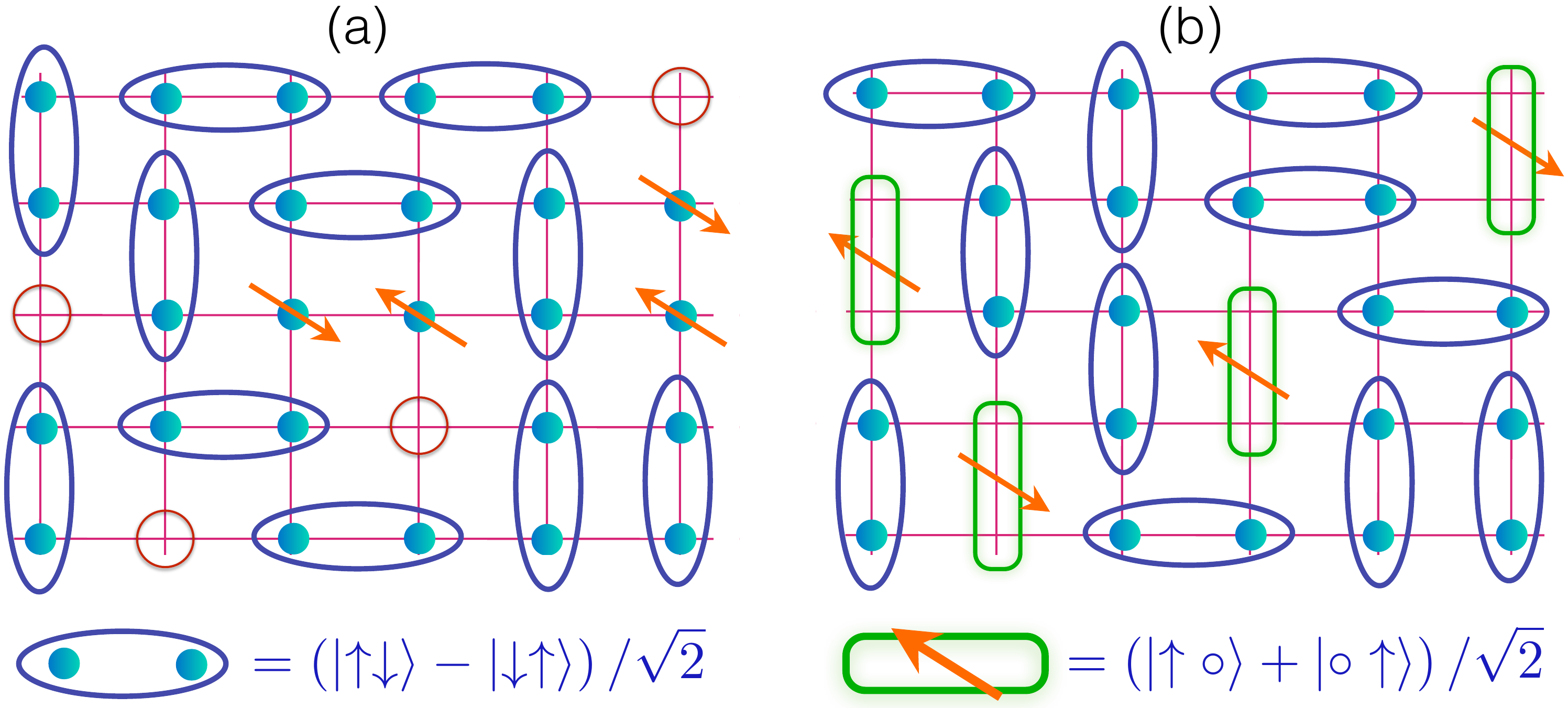}
\end{center}
\caption{(a) State obtained by breaking density $p/2$ valence bonds in Fig.~\ref{fig:holon}a into their constitute spinons.
(b) The spinons move into the neighborhood of the vacancies and form holon-spinon bound states represented by the green dimers \cite{Punk15}. 
The state with resonating blue and green dimers realizes a metal with a Fermi volume of $p$ quasiparticles with charge $+e$ and
spin $S=1/2$: the fractionalized Fermi liquid (FL*).}
\label{fig:fls}
\end{figure}

To obtain a spinful quasiparticle, we clearly have to attach an electronic spin to each holon. And as shown in Fig.~\ref{fig:fls}, 
it is not difficult to imagine
conditions under which this might be favorable: ({\em i\/}) We break density $p/2$ valence bonds into their constituent spins (Fig.~\ref{fig:fls}a); this costs some exchange energy for each valence bond broken. 
({\em ii\/}) We move the constituent spins (`spinons') into the neighborhood of the holons. ({\em iii\/}) The holons and spinons form a bound state (Fig.~\ref{fig:fls}b) 
which has both charge $+e$ and spin $S=1/2$, the same quantum numbers as (the absence of) an electron; this bound state formation gains energy which
can offset the energy cost of ({\em i\/}). 
We now have a modified resonating valence bond state \cite{Punk15}, like that in Eq.~(\ref{eq:rvb}), 
but with $\left| D_i \right\rangle$ consisting of pairing of sites of the square lattice with two categories of 
`valence bonds': the blue and green dimers in Fig.~\ref{fig:fls}b. 
The first class (blue) are the same as the electron singlet pairs
found in the Pauling-Anderson state. The second class (green) consists of a single electron resonating between the two sites
at the ends of the bond. From their constituents, it is clear that relative to the insulating RVB state, the blue dimers are
spinless, charge neutral bosons, while the green dimers are spin $S=1/2$, charge $+e$ fermions. 
Evidence that the states associated with the blue and green dimers dominate the wavefunction of the lightly-doped
cuprates appears in cluster dynamical mean field studies \cite{PG09,AMT11}.
Both classes of dimers
are mobile, and situation is somewhat analogous to $^{4}$He-$^{3}$He mixture. Like the $^{3}$He atoms, 
the green fermions can form a Fermi 
surface, and extension of the Luttinger argument to the present situation shows that the Fermi volume 
is exactly $p$ \cite{TSSSMV03,TSMVSS04,MPSS12}. However, unlike the $^{4}$He-$^{3}$He mixture, 
superfluidity is not immediate, because of the 
close-packing constraint on the blue+green dimers; onset of superfluidity will require pairing of the green dimers, and will not
be explored here. So the state obtained by resonating motion of the dimers in Fig.~\ref{fig:fls}b is a metal, dubbed
the fractionalized Fermi liquid (FL*) \cite{TSSSMV03}. It has a Fermi volume of $p$, with well-defined electron-like quasiparticles
near the Fermi surface.

A metallic state with a Fermi volume of density $p$ holes with charge $+e$ and spin $S=1/2$  
was initially described in Refs.~\cite{CS93,SS94}
by considering a theory for the loss of antiferromagnetic order in a doped antiferromagnetic state like that in 
Fig.~\ref{fig:fl}a. Numerous later studies \cite{XGWPAL96,XGW06,RKK07,RKK08,YQSS10,Mei12,MPSS12,Ferraz13,Punk15} 
described the resulting metallic state more completely in terms of the binding of holons and spinons, similar to the discussion above.
These studies also showed the presence of the emergent gauge excitations in the metal.

\begin{figure}
\begin{center}
\includegraphics[height=6.5cm]{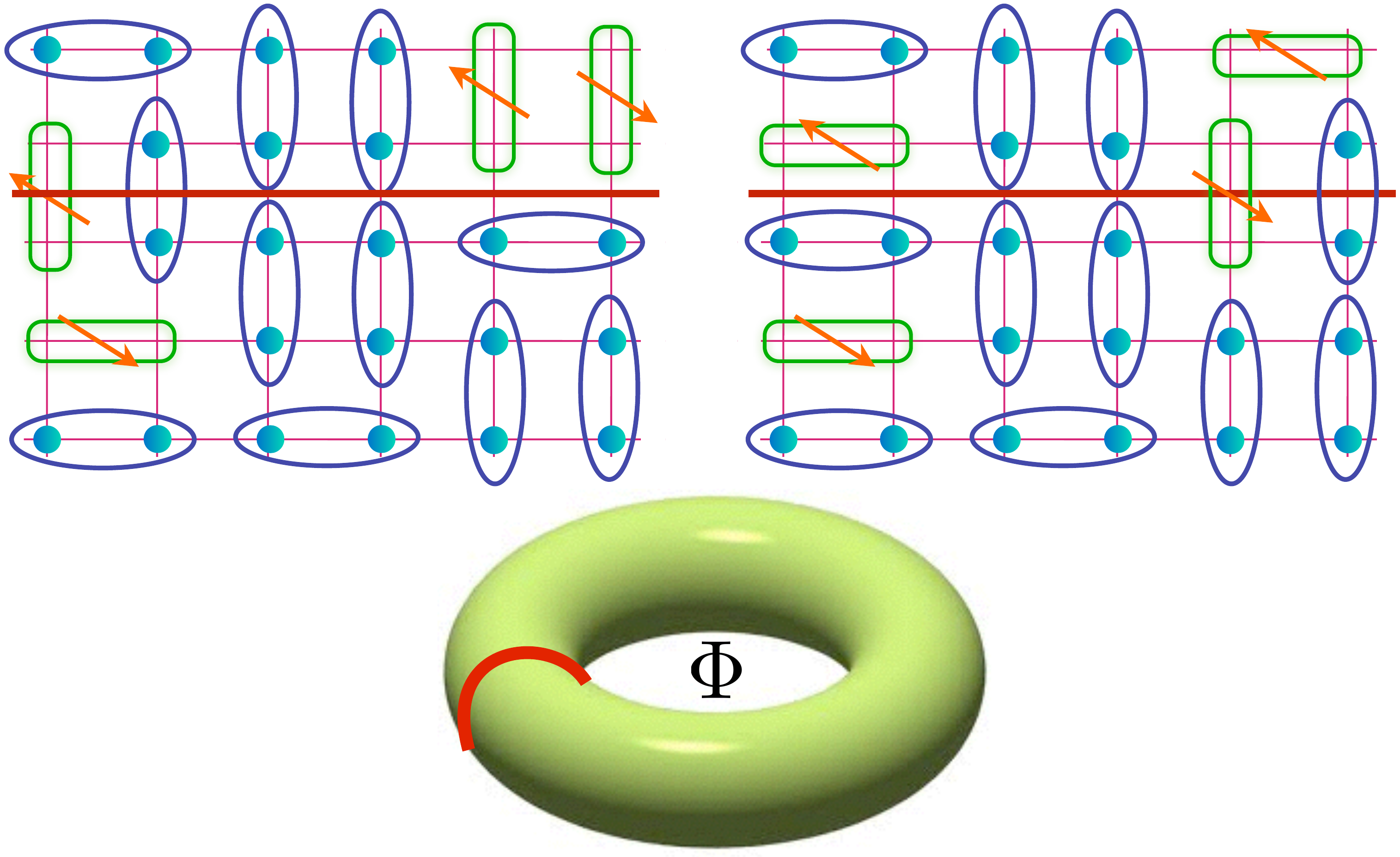}
\end{center}
\caption{Sample configurations of the wavefunction of the FL* state. The two configurations differ by local rearrangments
which preserve the sum of the number of blue {\em and\/} green dimers crossing the cut modulo 2, just as in Fig.~\ref{fig:torus} (only nearest-neighbor
dimers are shown for simplicity). This conservation implies the presence of gauge excitations, and additional states sensitive to the global topology, which cannot be decomposed into the quasiparticle excitations around the Fermi surface. Arguments in Refs.~\cite{MO00,TSMVSS04,APAV04}, 
based upon the adiabatic insertion of a fluxoid $\Phi = h/e$ through a cycle of the torus, show that the presence of these states allow the
Fermi volume to take a non-Luttinger value.}
\label{fig:torus2}
\end{figure}

Here, we can see the presence of emergent gauge fields, and associated low energy states
sensitive to the topology of the manifold, in our simplified description here of the FL* metal.
Indeed, such low energy states are required to evade the Luttinger theorem on the 
Fermi surface volume \cite{TSMVSS04,APAV04}. 
The FL* metal shares its topological features with corresponding insulating spin liquids, 
and we can transfer all of the arguments of Section~\ref{sec:sl}
practically unchanged, merely by applying them to wavefunctions like Fig.~\ref{fig:fls}b in a `color blind' manner.
So the arguments in Fig.~\ref{fig:torus} on the conservation of the number of valence bond across the cut modulo 2,
and associated near-degeneracies on the torus, apply equally to the FL* wavefunction after counting the numbers of {\em both\/}
blue and green dimers (see Fig.~\ref{fig:torus2}).
The presence of these near-degenerate topological states is also crucial for the Luttinger-volume-violating Fermi surface.
Oshikawa \cite{MO00} presented a proof of the Luttinger volume in a Fermi liquid by considering the consequences of 
adiabatically inserting a fluxoid $\Phi = h/e$ through a cycle of the torus, while assuming that the only low energy excitations on the torus are the quasiparticles around the Fermi surface. However, with the availibility of the low energy 
topological states discussed in Fig.~\ref{fig:torus2}, which are not related to quasiparticle excitations, it is possible to modify
Oshikawa's proof, and obtain a Fermi volume different from the Luttinger volume \cite{TSMVSS04,APAV04}; indeed a Luttinger volume 
of $p$ holes appears naturally in many models, including the simple models discussed here.

To summarize, this section has presented a simple description of a metallic state with the following features:
\begin{itemize}
\item A Fermi surface of holes of charge $e$ and spin $S=1/2$ enclosing volume $p$, and not the Luttinger volume of $1+p$.
\item Additional low energy quantum states on a torus not associated with quasiparticle excitations {\em i.e.\/} emergent gauge fields.
\end{itemize}
The flux-piercing arguments in Refs.~\cite{TSMVSS04,APAV04} show that it is not possible to have the first
feature without the second.

\section{The pseudogap metal of the cuprates}
\label{sec:pg}

\begin{figure}
\begin{center}
\includegraphics[height=5cm]{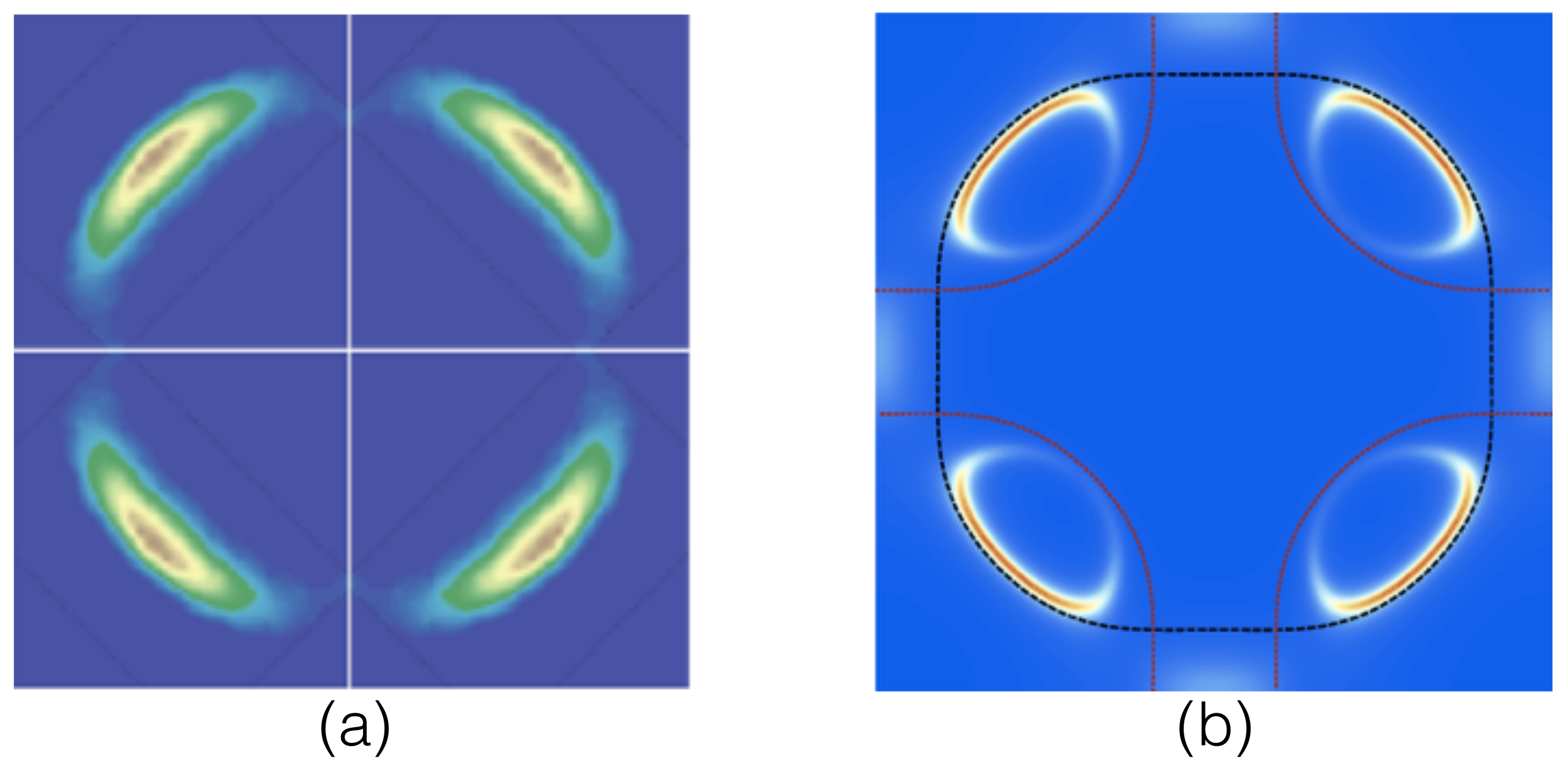}
\end{center}
\caption{(a) Photoemission spectrum in the PG regime \cite{Shen05}. Shown is the first Brillouin zone of the square lattice
centered at $(\pi, \pi)$, as in Fig.~\ref{fig:fl}c. However, unlike the FL result in Fig.~\ref{fig:fl}c, there is no continuous Fermi surface,
only `Fermi arcs'. (b) Photoemission from a FL* model in Ref.~\cite{YQSS10}. The Fermi surface consists of hole pockets of total area $p$, with intensity enhanced along the observed Fermi arcs, but suppressed on the `back' sides of the pockets.} 
\label{fig:photo}
\end{figure}

An early indication of the mysterious nature of the PG regime in Fig.~\ref{fig:phasediag}b was its
remarkable photoemission spectrum \cite{Shen05,Johnson10} (Fig.~\ref{fig:photo}).
There are low energy electronic excitations along the `nodal' directions in the form of `Fermi arcs', 
but none in the anti-nodal directions. In the FL* proposal, these arcs are remnants of hole pockets centered on the Brillouin
zone diagonals, with intensity suppressed on the `back' sides of the pockets \cite{YRZ,YQSS10}, as shown in Fig.~\ref{fig:photo}. 
The complete hole pockets have not
been observed in photoemission, but this is possibly accounted for by thermal broadening and weak intensity on their `back' sides.

More persuasive evidence for the FL* interpretation of the PG phase has come from a number of other experiments:
\begin{itemize}
\item A $T$-independent positive Hall co-efficient $R_H$ corresponding to carrier density $p$ in the higher temperature pseudogap \cite{Ando04}. This is the expected Hall co-efficient of the hole pockets in the FL* phase.
\item The frequency and temperature dependence of the optical conductivity has a Fermi liquid form $\sim 1/(-i \omega + 1/\tau)$ with $1/\tau \sim \omega^2 + T^2$ \cite{Marel13}. This Fermi liquid form is present although the overall prefactor corresponds
to a carrier density $p$.
\item Magnetoresistance measurements obey Kohler's rule \cite{MG14} with $\rho_{xx} \sim \tau^{-1} \left( 1 + a H^2 T^2 \right)$, again as expected by Fermi pocket of long-lived quasiparticles.
\item Density wave modulations have long been observed in STM experiments \cite{Kohsaka07} 
in the region marked DW in Fig.~\ref{fig:phasediag}b. Following theoretical proposals \cite{MMSS10b,SSRLP13}, a number of experiments \cite{Fujita14,Comin14,Forgan15,MHH15a,MHH15b} have identified the pattern of modulations as a $d$-form factor density wave. Computations of density wave instabilities of the FL* metal lead naturally to a $d$-form factor
density wave, with a wavevector similar to that observed in experiments \cite{DCSS14b}.
\item Finally, very interesting recent measurements by Badoux {\em et al.} \cite{Badoux15} 
of the Hall co-efficient at high fields and low $T$ for 
$p \approx 0.16$ in YBCO clearly show the absence of DW order, unlike those at lower $p$. Furthermore unlike the DW
region, the Hall co-efficient remains positive and corresponds to a density of $p$ carriers. Only at higher $p \approx 0.19$
does the FL Hall co-efficient of $1+p$ appear. A possible explanation is that the FL* phase is present
in the doping regime $0.16 < p < 0.19$ without the appearance of DW order. In Fig.~\ref{fig:phasediag}b, this corresponds
to the $T^\ast$ boundary extending past the DW region at low $T$.
\end{itemize}

\section{Conclusion}

We have described here the striking difference between the metallic states at low and high hole density, $p$,
in the cuprate superconductors (see Fig.~\ref{fig:phasediag}b). A theory for these states, and for the 
crossover between them, is clearly a needed precursor to any quantitative understanding of the high value of $T_c$
for the onset of superconductivity.

At high $p$, there is strong evidence for a conventional Fermi liquid (FL) state. This is an `unentangled' state, 
and its wavefunction is adiabatically connected to the free electron state which is a product of single-particle
Bloch waves. The Fermi surface has long-lived fermionic excitations with charge $+e$ and spin $S=1/2$.
The volume enclosed by the Fermi surface is $1+p$, and this obeys Luttinger's theorem.
Such a Fermi surface is seen clearly in photoemission experiments \cite{Dama05}, and also by the value
of the Hall co-efficient \cite{Badoux15}.

At low $p$, in the PG regime, the experimental results pose many interesting puzzles. Numerous transport measurements
\cite{Ando04,Marel13,MG14}, and also the remarkable 
recent Hall co-efficient measurements of Badoux {\em et al.\/} \cite{Badoux15}
at low $T$ and $p \approx 0.16$, are consistent with the presence of a Fermi liquid, but with a Fermi volume
of $p$, which is not the Luttinger value. We described basic aspects of the theory of the fractionalized Fermi liquid (FL*)
which realizes just such a Fermi surface. Long-range quantum entanglement, and emergent gauge fields, are
necessary ingredients which allow the FL* metal to have a Fermi surface enclosing a non-Luttinger 
volume. The FL* metal also leads to a possible understanding \cite{DCSS14b} of the density wave (DW) order found at low
temperatures in the pseudogap regime \cite{Fujita14,Comin14,Forgan15,MHH15a,MHH15b}.

Assuming the presence of the distinct FL and FL* metals at high and low $p$, we are faced with the central
open problem of connecting them at intermediate $p$. Although neither metal has a broken symmetry,
the presence of emergent gauge fields in the FL* implies that there cannot be an adiabatic connection
between the FL* and FL phases at zero temperature. So a quantum phase transition must be present,
but it is not in the Landau-Ginzburg-Wilson symmetry-breaking class. We need a quantum critical theory
with emergent gauge fields for the FL-FL* transition, and this can possibly provide an explanation for the
intermediate `strange metal' (SM) noted in Fig.~\ref{fig:phasediag}b. Examples of FL-FL* critical theories
have been proposed \cite{TSMVSS04,DCSS15b}, but a deeper understanding of such theories, and their connections to experimental
observations in the SM, remain important challenges for future research.

\enlargethispage{0pt}

%\ethics{Insert ethics statement here if applicable.}

%\dataccess{Insert details of how to access any supporting data here.}

%\aucontribute{For manuscripts with two or more authors, insert details of the authors contributions here. This should take the form: 'AB caried out the experiments. CD performed the data analysis. EF conceived of and designed the study, and drafted the manuscript All auhtors read and approved the manuscript'.}

\competing{The author declares that he has no competing interests.}

\funding{This research
was supported by the NSF under Grant DMR-1360789.
Research at Perimeter Institute is supported by the
Government of Canada through Industry Canada and by the Province of
Ontario through the Ministry of Research and Innovation. }

\ack{I thank Andrea Allais, Debanjan Chowdhury, S\'eamus Davis, Kazu Fujita, Antoine Georges, Mohammad Hamidian, Cyril Proust, Matthias Punk, 
and Louis Taillefer for numerous fruitful discussions on theories, experiments, and their connections.}

%\disclaimer{Insert disclaimer text here if applicable.}

%%%%%%%%%% Insert bibliography here %%%%%%%%%%%%%%

\bibliography{emergent}

\begin{thebibliography}{72}%
\makeatletter
\providecommand \@ifxundefined [1]{%
 \@ifx{#1\undefined}
}%
\providecommand \@ifnum [1]{%
 \ifnum #1\expandafter \@firstoftwo
 \else \expandafter \@secondoftwo
 \fi
}%
\providecommand \@ifx [1]{%
 \ifx #1\expandafter \@firstoftwo
 \else \expandafter \@secondoftwo
 \fi
}%
\providecommand \natexlab [1]{#1}%
\providecommand \enquote  [1]{``#1''}%
\providecommand \bibnamefont  [1]{#1}%
\providecommand \bibfnamefont [1]{#1}%
\providecommand \citenamefont [1]{#1}%
\providecommand \href@noop [0]{\@secondoftwo}%
\providecommand \href [0]{\begingroup \@sanitize@url \@href}%
\providecommand \@href[1]{\@@startlink{#1}\@@href}%
\providecommand \@@href[1]{\endgroup#1\@@endlink}%
\providecommand \@sanitize@url [0]{\catcode `\\12\catcode `\$12\catcode
  `\&12\catcode `\#12\catcode `\^12\catcode `\_12\catcode `\%12\relax}%
\providecommand \@@startlink[1]{}%
\providecommand \@@endlink[0]{}%
\providecommand \url  [0]{\begingroup\@sanitize@url \@url }%
\providecommand \@url [1]{\endgroup\@href {#1}{\urlprefix }}%
\providecommand \urlprefix  [0]{URL }%
\providecommand \Eprint [0]{\href }%
\providecommand \doibase [0]{http://dx.doi.org/}%
\providecommand \selectlanguage [0]{\@gobble}%
\providecommand \bibinfo  [0]{\@secondoftwo}%
\providecommand \bibfield  [0]{\@secondoftwo}%
\providecommand \translation [1]{[#1]}%
\providecommand \BibitemOpen [0]{}%
\providecommand \bibitemStop [0]{}%
\providecommand \bibitemNoStop [0]{.\EOS\space}%
\providecommand \EOS [0]{\spacefactor3000\relax}%
\providecommand \BibitemShut  [1]{\csname bibitem#1\endcsname}%
\let\auto@bib@innerbib\@empty
%</preamble>
\bibitem [{\citenamefont {{Mirzaei}}\ \emph {et~al.}(2013)\citenamefont
  {{Mirzaei}}, \citenamefont {{Stricker}}, \citenamefont {{Hancock}},
  \citenamefont {{Berthod}}, \citenamefont {{Georges}}, \citenamefont {{van
  Heumen}}, \citenamefont {{Chan}}, \citenamefont {{Zhao}}, \citenamefont
  {{Li}}, \citenamefont {{Greven}}, \citenamefont {{Barisic}},\ and\
  \citenamefont {{van der Marel}}}]{Marel13}%
  \BibitemOpen
  \bibfield  {author} {\bibinfo {author} {\bibfnamefont {S.~I.}\ \bibnamefont
  {{Mirzaei}}}, \bibinfo {author} {\bibfnamefont {D.}~\bibnamefont
  {{Stricker}}}, \bibinfo {author} {\bibfnamefont {J.~N.}\ \bibnamefont
  {{Hancock}}}, \bibinfo {author} {\bibfnamefont {C.}~\bibnamefont
  {{Berthod}}}, \bibinfo {author} {\bibfnamefont {A.}~\bibnamefont
  {{Georges}}}, \bibinfo {author} {\bibfnamefont {E.}~\bibnamefont {{van
  Heumen}}}, \bibinfo {author} {\bibfnamefont {M.~K.}\ \bibnamefont {{Chan}}},
  \bibinfo {author} {\bibfnamefont {X.}~\bibnamefont {{Zhao}}}, \bibinfo
  {author} {\bibfnamefont {Y.}~\bibnamefont {{Li}}}, \bibinfo {author}
  {\bibfnamefont {M.}~\bibnamefont {{Greven}}}, \bibinfo {author}
  {\bibfnamefont {N.}~\bibnamefont {{Barisic}}}, \ and\ \bibinfo {author}
  {\bibfnamefont {D.}~\bibnamefont {{van der Marel}}},\ }\bibfield  {title}
  {\enquote {\bibinfo {title} {{Spectroscopic evidence for Fermi liquid-like
  energy and temperature dependence of the relaxation rate in the pseudogap
  phase of the cuprates}},}\ }\href {\doibase 10.1073/pnas.1218846110}
  {\bibfield  {journal} {\bibinfo  {journal} {Proc. Nat. Acad. Sci.}\ }\textbf
  {\bibinfo {volume} {110}},\ \bibinfo {pages} {5774} (\bibinfo {year}
  {2013})},\ \Eprint {http://arxiv.org/abs/1207.6704}{arXiv:1207.6704
  [cond-mat.supr-con]}\BibitemShut {NoStop}%
\bibitem [{\citenamefont {Chan}\ \emph {et~al.}(2014)\citenamefont {Chan},
  \citenamefont {Veit}, \citenamefont {Dorow}, \citenamefont {Ge},
  \citenamefont {Li}, \citenamefont {Tabis}, \citenamefont {Tang},
  \citenamefont {Zhao}, \citenamefont {Bari\ifmmode \check{s}\else
  \v{s}\fi{}i\ifmmode~\acute{c}\else \'{c}\fi{}},\ and\ \citenamefont
  {Greven}}]{MG14}%
  \BibitemOpen
  \bibfield  {author} {\bibinfo {author} {\bibfnamefont {M.~K.}\ \bibnamefont
  {Chan}}, \bibinfo {author} {\bibfnamefont {M.~J.}\ \bibnamefont {Veit}},
  \bibinfo {author} {\bibfnamefont {C.~J.}\ \bibnamefont {Dorow}}, \bibinfo
  {author} {\bibfnamefont {Y.}~\bibnamefont {Ge}}, \bibinfo {author}
  {\bibfnamefont {Y.}~\bibnamefont {Li}}, \bibinfo {author} {\bibfnamefont
  {W.}~\bibnamefont {Tabis}}, \bibinfo {author} {\bibfnamefont
  {Y.}~\bibnamefont {Tang}}, \bibinfo {author} {\bibfnamefont {X.}~\bibnamefont
  {Zhao}}, \bibinfo {author} {\bibfnamefont {N.}~\bibnamefont {Bari\ifmmode
  \check{s}\else \v{s}\fi{}i\ifmmode~\acute{c}\else \'{c}\fi{}}}, \ and\
  \bibinfo {author} {\bibfnamefont {M.}~\bibnamefont {Greven}},\ }\bibfield
  {title} {\enquote {\bibinfo {title} {{In-Plane Magnetoresistance Obeys
  Kohler's Rule in the Pseudogap Phase of Cuprate Superconductors}},}\ }\href
  {\doibase 10.1103/PhysRevLett.113.177005} {\bibfield  {journal} {\bibinfo
  {journal} {Phys. Rev. Lett.}\ }\textbf {\bibinfo {volume} {113}},\ \bibinfo
  {pages} {177005} (\bibinfo {year} {2014})},\ \Eprint
  {http://arxiv.org/abs/1402.4472}{arXiv:1402.4472
  [cond-mat.supr-con]}\BibitemShut {NoStop}%
\bibitem [{\citenamefont {{Fujita}}\ \emph {et~al.}(2014)\citenamefont
  {{Fujita}}, \citenamefont {{Hamidian}}, \citenamefont {{Edkins}},
  \citenamefont {{Kim}}, \citenamefont {{Kohsaka}}, \citenamefont {{Azuma}},
  \citenamefont {{Takano}}, \citenamefont {{Takagi}}, \citenamefont {{Eisaki}},
  \citenamefont {{Uchida}}, \citenamefont {{Allais}}, \citenamefont {{Lawler}},
  \citenamefont {{Kim}}, \citenamefont {{Sachdev}},\ and\ \citenamefont
  {{S{\'e}amus Davis}}}]{Fujita14}%
  \BibitemOpen
  \bibfield  {author} {\bibinfo {author} {\bibfnamefont {K.}~\bibnamefont
  {{Fujita}}}, \bibinfo {author} {\bibfnamefont {M.~H.}\ \bibnamefont
  {{Hamidian}}}, \bibinfo {author} {\bibfnamefont {S.~D.}\ \bibnamefont
  {{Edkins}}}, \bibinfo {author} {\bibfnamefont {C.~K.}\ \bibnamefont {{Kim}}},
  \bibinfo {author} {\bibfnamefont {Y.}~\bibnamefont {{Kohsaka}}}, \bibinfo
  {author} {\bibfnamefont {M.}~\bibnamefont {{Azuma}}}, \bibinfo {author}
  {\bibfnamefont {M.}~\bibnamefont {{Takano}}}, \bibinfo {author}
  {\bibfnamefont {H.}~\bibnamefont {{Takagi}}}, \bibinfo {author}
  {\bibfnamefont {H.}~\bibnamefont {{Eisaki}}}, \bibinfo {author}
  {\bibfnamefont {S.-i.}\ \bibnamefont {{Uchida}}}, \bibinfo {author}
  {\bibfnamefont {A.}~\bibnamefont {{Allais}}}, \bibinfo {author}
  {\bibfnamefont {M.~J.}\ \bibnamefont {{Lawler}}}, \bibinfo {author}
  {\bibfnamefont {E.-A.}\ \bibnamefont {{Kim}}}, \bibinfo {author}
  {\bibfnamefont {S.}~\bibnamefont {{Sachdev}}}, \ and\ \bibinfo {author}
  {\bibfnamefont {J.~C.}\ \bibnamefont {{S{\'e}amus Davis}}},\ }\bibfield
  {title} {\enquote {\bibinfo {title} {{Direct phase-sensitive identification
  of a d-form factor density wave in underdoped cuprates}},}\ }\href {\doibase
  10.1073/pnas.1406297111} {\bibfield  {journal} {\bibinfo  {journal} {Proc.
  Nat. Acad. Sci.}\ }\textbf {\bibinfo {volume} {111}},\ \bibinfo {pages}
  {3026} (\bibinfo {year} {2014})},\ \Eprint
  {http://arxiv.org/abs/1404.0362}{arXiv:1404.0362
  [cond-mat.supr-con]}\BibitemShut {NoStop}%
\bibitem [{\citenamefont {{Comin}}\ \emph {et~al.}(2015)\citenamefont
  {{Comin}}, \citenamefont {{Sutarto}}, \citenamefont {{He}}, \citenamefont
  {{da Silva Neto}}, \citenamefont {{Chauviere}}, \citenamefont {{Fra{\~n}o}},
  \citenamefont {{Liang}}, \citenamefont {{Hardy}}, \citenamefont {{Bonn}},
  \citenamefont {{Yoshida}}, \citenamefont {{Eisaki}}, \citenamefont
  {{Achkar}}, \citenamefont {{Hawthorn}}, \citenamefont {{Keimer}},
  \citenamefont {{Sawatzky}},\ and\ \citenamefont {{Damascelli}}}]{Comin14}%
  \BibitemOpen
  \bibfield  {author} {\bibinfo {author} {\bibfnamefont {R.}~\bibnamefont
  {{Comin}}}, \bibinfo {author} {\bibfnamefont {R.}~\bibnamefont {{Sutarto}}},
  \bibinfo {author} {\bibfnamefont {F.}~\bibnamefont {{He}}}, \bibinfo {author}
  {\bibfnamefont {E.~H.}\ \bibnamefont {{da Silva Neto}}}, \bibinfo {author}
  {\bibfnamefont {L.}~\bibnamefont {{Chauviere}}}, \bibinfo {author}
  {\bibfnamefont {A.}~\bibnamefont {{Fra{\~n}o}}}, \bibinfo {author}
  {\bibfnamefont {R.}~\bibnamefont {{Liang}}}, \bibinfo {author} {\bibfnamefont
  {W.~N.}\ \bibnamefont {{Hardy}}}, \bibinfo {author} {\bibfnamefont {D.~A.}\
  \bibnamefont {{Bonn}}}, \bibinfo {author} {\bibfnamefont {Y.}~\bibnamefont
  {{Yoshida}}}, \bibinfo {author} {\bibfnamefont {H.}~\bibnamefont {{Eisaki}}},
  \bibinfo {author} {\bibfnamefont {A.~J.}\ \bibnamefont {{Achkar}}}, \bibinfo
  {author} {\bibfnamefont {D.~G.}\ \bibnamefont {{Hawthorn}}}, \bibinfo
  {author} {\bibfnamefont {B.}~\bibnamefont {{Keimer}}}, \bibinfo {author}
  {\bibfnamefont {G.~A.}\ \bibnamefont {{Sawatzky}}}, \ and\ \bibinfo {author}
  {\bibfnamefont {A.}~\bibnamefont {{Damascelli}}},\ }\bibfield  {title}
  {\enquote {\bibinfo {title} {{Symmetry of charge order in cuprates}},}\
  }\href {\doibase 10.1038/nmat4295} {\bibfield  {journal} {\bibinfo  {journal}
  {Nature Materials}\ }\textbf {\bibinfo {volume} {14}},\ \bibinfo {pages}
  {796} (\bibinfo {year} {2015})},\ \Eprint
  {http://arxiv.org/abs/1402.5415}{arXiv:1402.5415
  [cond-mat.supr-con]}\BibitemShut {NoStop}%
\bibitem [{\citenamefont {{Forgan}}\ \emph {et~al.}(2015)\citenamefont
  {{Forgan}}, \citenamefont {{Blackburn}}, \citenamefont {{Holmes}},
  \citenamefont {{Briffa}}, \citenamefont {{Chang}}, \citenamefont
  {{Bouchenoire}}, \citenamefont {{Brown}}, \citenamefont {{Liang}},
  \citenamefont {{Bonn}}, \citenamefont {{Hardy}}, \citenamefont
  {{Christensen}}, \citenamefont {{Zimmermann}}, \citenamefont {{Huecker}},\
  and\ \citenamefont {{Hayden}}}]{Forgan15}%
  \BibitemOpen
  \bibfield  {author} {\bibinfo {author} {\bibfnamefont {E.~M.}\ \bibnamefont
  {{Forgan}}}, \bibinfo {author} {\bibfnamefont {E.}~\bibnamefont
  {{Blackburn}}}, \bibinfo {author} {\bibfnamefont {A.~T.}\ \bibnamefont
  {{Holmes}}}, \bibinfo {author} {\bibfnamefont {A.}~\bibnamefont {{Briffa}}},
  \bibinfo {author} {\bibfnamefont {J.}~\bibnamefont {{Chang}}}, \bibinfo
  {author} {\bibfnamefont {L.}~\bibnamefont {{Bouchenoire}}}, \bibinfo {author}
  {\bibfnamefont {S.~D.}\ \bibnamefont {{Brown}}}, \bibinfo {author}
  {\bibfnamefont {R.}~\bibnamefont {{Liang}}}, \bibinfo {author} {\bibfnamefont
  {D.}~\bibnamefont {{Bonn}}}, \bibinfo {author} {\bibfnamefont {W.~N.}\
  \bibnamefont {{Hardy}}}, \bibinfo {author} {\bibfnamefont {N.~B.}\
  \bibnamefont {{Christensen}}}, \bibinfo {author} {\bibfnamefont {M.~v.}\
  \bibnamefont {{Zimmermann}}}, \bibinfo {author} {\bibfnamefont
  {M.}~\bibnamefont {{Huecker}}}, \ and\ \bibinfo {author} {\bibfnamefont
  {S.~M.}\ \bibnamefont {{Hayden}}},\ }\bibfield  {title} {\enquote {\bibinfo
  {title} {{The nature of the charge density waves in under-doped
  YBa$_2$Cu$_3$O$_{6.54}$ revealed by X-ray measurements of the ionic
  displacements}},}\ }\href@noop {} {\bibfield  {journal} {\bibinfo  {journal}
  {ArXiv e-prints}\ } (\bibinfo {year} {2015})},\ \Eprint
  {http://arxiv.org/abs/1504.01585}{arXiv:1504.01585
  [cond-mat.supr-con]}\BibitemShut {NoStop}%
\bibitem [{\citenamefont {Hamidian}\ \emph {et~al.}(2015)\citenamefont
  {Hamidian}, \citenamefont {Edkins}, \citenamefont {Kim}, \citenamefont
  {{S{\'e}amus Davis}}, \citenamefont {Mackenzie}, \citenamefont {Eisaki},
  \citenamefont {Uchida}, \citenamefont {Lawler}, \citenamefont {Kim},
  \citenamefont {Sachdev},\ and\ \citenamefont {Fujita}}]{MHH15a}%
  \BibitemOpen
  \bibfield  {author} {\bibinfo {author} {\bibfnamefont {M.~H.}\ \bibnamefont
  {Hamidian}}, \bibinfo {author} {\bibfnamefont {S.~D.}\ \bibnamefont
  {Edkins}}, \bibinfo {author} {\bibfnamefont {C.~K.}\ \bibnamefont {Kim}},
  \bibinfo {author} {\bibfnamefont {J.~C.}\ \bibnamefont {{S{\'e}amus Davis}}},
  \bibinfo {author} {\bibfnamefont {A.~P.}\ \bibnamefont {Mackenzie}}, \bibinfo
  {author} {\bibfnamefont {H.}~\bibnamefont {Eisaki}}, \bibinfo {author}
  {\bibfnamefont {S.}~\bibnamefont {Uchida}}, \bibinfo {author} {\bibfnamefont
  {M.~J.}\ \bibnamefont {Lawler}}, \bibinfo {author} {\bibfnamefont {E.~A.}\
  \bibnamefont {Kim}}, \bibinfo {author} {\bibfnamefont {S.}~\bibnamefont
  {Sachdev}}, \ and\ \bibinfo {author} {\bibfnamefont {K.}~\bibnamefont
  {Fujita}},\ }\bibfield  {title} {\enquote {\bibinfo {title} {{Atomic-scale
  electronic structure of the cuprate $d$-symmetry form factor density wave
  state}},}\ }\href {http://dx.doi.org/10.1038/nphys3519} {\bibfield  {journal}
  {\bibinfo  {journal} {Nature Physics}\ }\textbf {\bibinfo {volume} {advance
  online publication}},\  (\bibinfo {year} {2015})},\ \Eprint
  {http://arxiv.org/abs/1507.07865}{arXiv:1507.07865
  [cond-mat.supr-con]}\BibitemShut {NoStop}%
\bibitem [{\citenamefont {{Hamidian}}\ \emph {et~al.}(2015)\citenamefont
  {{Hamidian}}, \citenamefont {{Edkins}}, \citenamefont {{Fujita}},
  \citenamefont {{Kostin}}, \citenamefont {{Mackenzie}}, \citenamefont
  {{Eisaki}}, \citenamefont {{Uchida}}, \citenamefont {{Lawler}}, \citenamefont
  {{Kim}}, \citenamefont {{Sachdev}},\ and\ \citenamefont {{S{\'e}amus
  Davis}}}]{MHH15b}%
  \BibitemOpen
  \bibfield  {author} {\bibinfo {author} {\bibfnamefont {M.~H.}\ \bibnamefont
  {{Hamidian}}}, \bibinfo {author} {\bibfnamefont {S.~D.}\ \bibnamefont
  {{Edkins}}}, \bibinfo {author} {\bibfnamefont {K.}~\bibnamefont {{Fujita}}},
  \bibinfo {author} {\bibfnamefont {A.}~\bibnamefont {{Kostin}}}, \bibinfo
  {author} {\bibfnamefont {A.~P.}\ \bibnamefont {{Mackenzie}}}, \bibinfo
  {author} {\bibfnamefont {H.}~\bibnamefont {{Eisaki}}}, \bibinfo {author}
  {\bibfnamefont {S.}~\bibnamefont {{Uchida}}}, \bibinfo {author}
  {\bibfnamefont {M.~J.}\ \bibnamefont {{Lawler}}}, \bibinfo {author}
  {\bibfnamefont {E.-A.}\ \bibnamefont {{Kim}}}, \bibinfo {author}
  {\bibfnamefont {S.}~\bibnamefont {{Sachdev}}}, \ and\ \bibinfo {author}
  {\bibfnamefont {J.~C.}\ \bibnamefont {{S{\'e}amus Davis}}},\ }\bibfield
  {title} {\enquote {\bibinfo {title} {{Magnetic-field Induced Interconversion
  of Cooper Pairs and Density Wave States within Cuprate Composite Order}},}\
  }\href@noop {} {\bibfield  {journal} {\bibinfo  {journal} {ArXiv e-prints}\ }
  (\bibinfo {year} {2015})},\ \Eprint
  {http://arxiv.org/abs/1508.00620}{arXiv:1508.00620
  [cond-mat.supr-con]}\BibitemShut {NoStop}%
\bibitem [{\citenamefont {{Badoux}}\ \emph {et~al.}(2015)\citenamefont
  {{Badoux}}, \citenamefont {{Tabis}}, \citenamefont {{Lalibert{\'e}}},
  \citenamefont {{Grissonnanche}}, \citenamefont {{Vignolle}}, \citenamefont
  {{Vignolles}}, \citenamefont {{B{\'e}ard}}, \citenamefont {{Bonn}},
  \citenamefont {{Hardy}}, \citenamefont {{Liang}}, \citenamefont
  {{Doiron-Leyraud}}, \citenamefont {{Taillefer}},\ and\ \citenamefont
  {{Proust}}}]{Badoux15}%
  \BibitemOpen
  \bibfield  {author} {\bibinfo {author} {\bibfnamefont {S.}~\bibnamefont
  {{Badoux}}}, \bibinfo {author} {\bibfnamefont {W.}~\bibnamefont {{Tabis}}},
  \bibinfo {author} {\bibfnamefont {F.}~\bibnamefont {{Lalibert{\'e}}}},
  \bibinfo {author} {\bibfnamefont {G.}~\bibnamefont {{Grissonnanche}}},
  \bibinfo {author} {\bibfnamefont {B.}~\bibnamefont {{Vignolle}}}, \bibinfo
  {author} {\bibfnamefont {D.}~\bibnamefont {{Vignolles}}}, \bibinfo {author}
  {\bibfnamefont {J.}~\bibnamefont {{B{\'e}ard}}}, \bibinfo {author}
  {\bibfnamefont {D.~A.}\ \bibnamefont {{Bonn}}}, \bibinfo {author}
  {\bibfnamefont {W.~N.}\ \bibnamefont {{Hardy}}}, \bibinfo {author}
  {\bibfnamefont {R.}~\bibnamefont {{Liang}}}, \bibinfo {author} {\bibfnamefont
  {N.}~\bibnamefont {{Doiron-Leyraud}}}, \bibinfo {author} {\bibfnamefont
  {L.}~\bibnamefont {{Taillefer}}}, \ and\ \bibinfo {author} {\bibfnamefont
  {C.}~\bibnamefont {{Proust}}},\ }\bibfield  {title} {\enquote {\bibinfo
  {title} {{Change of carrier density at the pseudogap critical point of a
  cuprate superconductor}},}\ }\href@noop {} {\bibfield  {journal} {\bibinfo
  {journal} {ArXiv e-prints}\ } (\bibinfo {year} {2015})},\ \Eprint
  {http://arxiv.org/abs/1511.08162}{arXiv:1511.08162
  [cond-mat.supr-con]}\BibitemShut {NoStop}%
\bibitem [{\citenamefont {Chowdhury}\ and\ \citenamefont
  {Sachdev}(2015)}]{DCSS15}%
  \BibitemOpen
  \bibfield  {author} {\bibinfo {author} {\bibfnamefont {D.}~\bibnamefont
  {Chowdhury}}\ and\ \bibinfo {author} {\bibfnamefont {S.}~\bibnamefont
  {Sachdev}},\ }\bibfield  {title} {\enquote {\bibinfo {title} {{The enigma of
  the pseudogap phase of the cuprate superconductors}},}\ }\href@noop {}
  {\bibfield  {journal} {\bibinfo  {journal} {arXiv:1501.00002}\ } (\bibinfo
  {year} {2015})},\ \Eprint {http://arxiv.org/abs/1501.00002}{arXiv:1501.00002
  [cond-mat.str-el]}\BibitemShut {NoStop}%
\bibitem [{\citenamefont {Sachdev}(2004)}]{SS04}%
  \BibitemOpen
  \bibfield  {author} {\bibinfo {author} {\bibfnamefont {S.}~\bibnamefont
  {Sachdev}},\ }\bibfield  {title} {\enquote {\bibinfo {title} {{Quantum phases
  and phase transitions of Mott insulators}},}\ }in\ \href {\doibase
  10.1007/BFb0119599} {\emph {\bibinfo {booktitle} {Quantum Magnetism}}},\
  \bibinfo {series} {Lecture Notes in Physics}, Vol.\ \bibinfo {volume} {645},\
  \bibinfo {editor} {edited by\ \bibinfo {editor} {\bibfnamefont
  {U.}~\bibnamefont {Schollw\"ock}}, \bibinfo {editor} {\bibfnamefont
  {J.}~\bibnamefont {Richter}}, \bibinfo {editor} {\bibfnamefont {D.~J.~J.}\
  \bibnamefont {Farnell}}, \ and\ \bibinfo {editor} {\bibfnamefont {R.~F.}\
  \bibnamefont {Bishop}}}\ (\bibinfo  {publisher} {Springer Berlin
  Heidelberg},\ \bibinfo {year} {2004})\ pp.\ \bibinfo {pages} {381--432},\
  \Eprint {http://arxiv.org/abs/cond-mat/0401041}{cond-mat/0401041}\BibitemShut
  {NoStop}%
\bibitem [{\citenamefont {Sachdev}(2012)}]{SS10}%
  \BibitemOpen
  \bibfield  {author} {\bibinfo {author} {\bibfnamefont {S.}~\bibnamefont
  {Sachdev}},\ }\bibfield  {title} {\enquote {\bibinfo {title} {Quantum phase
  transitions of antiferromagnets and the cuprate superconductors},}\ }in\
  \href {\doibase 10.1007/978-3-642-10449-7_1} {\emph {\bibinfo {booktitle}
  {{Modern Theories of Many-Particle Systems in Condensed Matter Physics}}}},\
  \bibinfo {series} {Lecture Notes in Physics}, Vol.\ \bibinfo {volume} {843,
  2009 Les Houches school},\ \bibinfo {editor} {edited by\ \bibinfo {editor}
  {\bibfnamefont {D.~C.}\ \bibnamefont {Cabra}}, \bibinfo {editor}
  {\bibfnamefont {A.}~\bibnamefont {Honecker}}, \ and\ \bibinfo {editor}
  {\bibfnamefont {P.}~\bibnamefont {Pujol}}}\ (\bibinfo  {publisher} {Springer
  Berlin Heidelberg},\ \bibinfo {year} {2012})\ pp.\ \bibinfo {pages} {1--51},\
  \Eprint {http://arxiv.org/abs/1002.3823}{arXiv:1002.3823
  [cond-mat.str-el]}\BibitemShut {NoStop}%
\bibitem [{\citenamefont {Pauling}(1949)}]{Pauling49}%
  \BibitemOpen
  \bibfield  {author} {\bibinfo {author} {\bibfnamefont {L.}~\bibnamefont
  {Pauling}},\ }\bibfield  {title} {\enquote {\bibinfo {title} {A
  resonating-valence-bond theory of metals and intermetallic compounds},}\
  }\href {\doibase 10.1098/rspa.1949.0032} {\bibfield  {journal} {\bibinfo
  {journal} {Proceedings of the Royal Society of London A: Mathematical,
  Physical and Engineering Sciences}\ }\textbf {\bibinfo {volume} {196}},\
  \bibinfo {pages} {343} (\bibinfo {year} {1949})}\BibitemShut {NoStop}%
\bibitem [{\citenamefont {Anderson}(1973)}]{Anderson73}%
  \BibitemOpen
  \bibfield  {author} {\bibinfo {author} {\bibfnamefont {P.~W.}\ \bibnamefont
  {Anderson}},\ }\bibfield  {title} {\enquote {\bibinfo {title} {Resonating
  valence bonds: A new kind of insulator?}}\ }\href {\doibase
  http://dx.doi.org/10.1016/0025-5408(73)90167-0} {\bibfield  {journal}
  {\bibinfo  {journal} {Materials Research Bulletin}\ }\textbf {\bibinfo
  {volume} {8}},\ \bibinfo {pages} {153 } (\bibinfo {year} {1973})}\BibitemShut
  {NoStop}%
\bibitem [{\citenamefont {Laughlin}(1983)}]{RBL83}%
  \BibitemOpen
  \bibfield  {author} {\bibinfo {author} {\bibfnamefont {R.~B.}\ \bibnamefont
  {Laughlin}},\ }\bibfield  {title} {\enquote {\bibinfo {title} {{Anomalous
  Quantum Hall Effect: An Incompressible Quantum Fluid with Fractionally
  Charged Excitations}},}\ }\href {\doibase 10.1103/PhysRevLett.50.1395}
  {\bibfield  {journal} {\bibinfo  {journal} {Phys. Rev. Lett.}\ }\textbf
  {\bibinfo {volume} {50}},\ \bibinfo {pages} {1395} (\bibinfo {year}
  {1983})}\BibitemShut {NoStop}%
\bibitem [{\citenamefont {Kalmeyer}\ and\ \citenamefont
  {Laughlin}(1987)}]{RBL87}%
  \BibitemOpen
  \bibfield  {author} {\bibinfo {author} {\bibfnamefont {V.}~\bibnamefont
  {Kalmeyer}}\ and\ \bibinfo {author} {\bibfnamefont {R.~B.}\ \bibnamefont
  {Laughlin}},\ }\bibfield  {title} {\enquote {\bibinfo {title} {{Equivalence
  of the resonating-valence-bond and fractional quantum Hall states}},}\ }\href
  {\doibase 10.1103/PhysRevLett.59.2095} {\bibfield  {journal} {\bibinfo
  {journal} {Phys. Rev. Lett.}\ }\textbf {\bibinfo {volume} {59}},\ \bibinfo
  {pages} {2095} (\bibinfo {year} {1987})}\BibitemShut {NoStop}%
\bibitem [{\citenamefont {{Kitaev}}\ and\ \citenamefont
  {{Preskill}}(2006)}]{AKJP05}%
  \BibitemOpen
  \bibfield  {author} {\bibinfo {author} {\bibfnamefont {A.}~\bibnamefont
  {{Kitaev}}}\ and\ \bibinfo {author} {\bibfnamefont {J.}~\bibnamefont
  {{Preskill}}},\ }\bibfield  {title} {\enquote {\bibinfo {title} {{Topological
  Entanglement Entropy}},}\ }\href {\doibase 10.1103/PhysRevLett.96.110404}
  {\bibfield  {journal} {\bibinfo  {journal} {Phys. Rev. Lett.}\ }\textbf
  {\bibinfo {volume} {96}},\ \bibinfo {eid} {110404} (\bibinfo {year}
  {2006})},\ \Eprint
  {http://arxiv.org/abs/hep-th/0510092}{hep-th/0510092}\BibitemShut {NoStop}%
\bibitem [{\citenamefont {{Levin}}\ and\ \citenamefont {{Wen}}(2006)}]{XGW05}%
  \BibitemOpen
  \bibfield  {author} {\bibinfo {author} {\bibfnamefont {M.}~\bibnamefont
  {{Levin}}}\ and\ \bibinfo {author} {\bibfnamefont {X.-G.}\ \bibnamefont
  {{Wen}}},\ }\bibfield  {title} {\enquote {\bibinfo {title} {{Detecting
  Topological Order in a Ground State Wave Function}},}\ }\href {\doibase
  10.1103/PhysRevLett.96.110405} {\bibfield  {journal} {\bibinfo  {journal}
  {Phys. Rev. Lett.}\ }\textbf {\bibinfo {volume} {96}},\ \bibinfo {eid}
  {110405} (\bibinfo {year} {2006})},\ \Eprint
  {http://arxiv.org/abs/cond-mat/0510613}{cond-mat/0510613}\BibitemShut
  {NoStop}%
\bibitem [{\citenamefont {{Wildeboer}}\ \emph {et~al.}(2015)\citenamefont
  {{Wildeboer}}, \citenamefont {{Seidel}},\ and\ \citenamefont
  {{Melko}}}]{Melko15}%
  \BibitemOpen
  \bibfield  {author} {\bibinfo {author} {\bibfnamefont {J.}~\bibnamefont
  {{Wildeboer}}}, \bibinfo {author} {\bibfnamefont {A.}~\bibnamefont
  {{Seidel}}}, \ and\ \bibinfo {author} {\bibfnamefont {R.~G.}\ \bibnamefont
  {{Melko}}},\ }\bibfield  {title} {\enquote {\bibinfo {title} {{Entanglement
  Entropy and Topological Order in Resonating Valence-Bond Quantum Spin
  Liquids}},}\ }\href@noop {} {\bibfield  {journal} {\bibinfo  {journal} {ArXiv
  e-prints}\ } (\bibinfo {year} {2015})},\ \Eprint
  {http://arxiv.org/abs/1510.07682}{arXiv:1510.07682
  [cond-mat.str-el]}\BibitemShut {NoStop}%
\bibitem [{\citenamefont {Thouless}(1987)}]{Thouless87}%
  \BibitemOpen
  \bibfield  {author} {\bibinfo {author} {\bibfnamefont {D.~J.}\ \bibnamefont
  {Thouless}},\ }\bibfield  {title} {\enquote {\bibinfo {title} {Fluxoid
  quantization in the resonating-valence-bond model},}\ }\href {\doibase
  10.1103/PhysRevB.36.7187} {\bibfield  {journal} {\bibinfo  {journal} {Phys.
  Rev. B}\ }\textbf {\bibinfo {volume} {36}},\ \bibinfo {pages} {7187}
  (\bibinfo {year} {1987})}\BibitemShut {NoStop}%
\bibitem [{\citenamefont {Kivelson}\ \emph {et~al.}(1988)\citenamefont
  {Kivelson}, \citenamefont {Rokhsar},\ and\ \citenamefont {Sethna}}]{KRS88}%
  \BibitemOpen
  \bibfield  {author} {\bibinfo {author} {\bibfnamefont {S.~A.}\ \bibnamefont
  {Kivelson}}, \bibinfo {author} {\bibfnamefont {D.~S.}\ \bibnamefont
  {Rokhsar}}, \ and\ \bibinfo {author} {\bibfnamefont {J.~P.}\ \bibnamefont
  {Sethna}},\ }\bibfield  {title} {\enquote {\bibinfo {title} {{$2e$ or not
  $2e$ : Flux Quantization in the Resonating Valence Bond State}},}\ }\href
  {http://stacks.iop.org/0295-5075/6/i=4/a=013} {\bibfield  {journal} {\bibinfo
   {journal} {Europhysics Letters}\ }\textbf {\bibinfo {volume} {6}},\ \bibinfo
  {pages} {353} (\bibinfo {year} {1988})}\BibitemShut {NoStop}%
\bibitem [{\citenamefont {Baskaran}\ and\ \citenamefont
  {Anderson}(1988)}]{GBPWA88}%
  \BibitemOpen
  \bibfield  {author} {\bibinfo {author} {\bibfnamefont {G.}~\bibnamefont
  {Baskaran}}\ and\ \bibinfo {author} {\bibfnamefont {P.~W.}\ \bibnamefont
  {Anderson}},\ }\bibfield  {title} {\enquote {\bibinfo {title} {{Gauge theory
  of high-temperature superconductors and strongly correlated Fermi
  systems}},}\ }\href {\doibase 10.1103/PhysRevB.37.580} {\bibfield  {journal}
  {\bibinfo  {journal} {Phys. Rev. B}\ }\textbf {\bibinfo {volume} {37}},\
  \bibinfo {pages} {580} (\bibinfo {year} {1988})}\BibitemShut {NoStop}%
\bibitem [{\citenamefont {Fradkin}\ and\ \citenamefont
  {Kivelson}(1990)}]{EFSK90}%
  \BibitemOpen
  \bibfield  {author} {\bibinfo {author} {\bibfnamefont {E.}~\bibnamefont
  {Fradkin}}\ and\ \bibinfo {author} {\bibfnamefont {S.~A.}\ \bibnamefont
  {Kivelson}},\ }\bibfield  {title} {\enquote {\bibinfo {title} {Short range
  resonating valence bond theories and superconductivity},}\ }\href {\doibase
  10.1142/S0217984990000295} {\bibfield  {journal} {\bibinfo  {journal} {Mod.
  Phys. Lett. B}\ }\textbf {\bibinfo {volume} {04}},\ \bibinfo {pages} {225}
  (\bibinfo {year} {1990})}\BibitemShut {NoStop}%
\bibitem [{\citenamefont {Read}\ and\ \citenamefont {Sachdev}(1989)}]{NRSS89}%
  \BibitemOpen
  \bibfield  {author} {\bibinfo {author} {\bibfnamefont {N.}~\bibnamefont
  {Read}}\ and\ \bibinfo {author} {\bibfnamefont {S.}~\bibnamefont {Sachdev}},\
  }\bibfield  {title} {\enquote {\bibinfo {title} {{Valence-bond and
  spin-Peierls ground states of low-dimensional quantum antiferromagnets}},}\
  }\href {\doibase 10.1103/PhysRevLett.62.1694} {\bibfield  {journal} {\bibinfo
   {journal} {Phys. Rev. Lett.}\ }\textbf {\bibinfo {volume} {62}},\ \bibinfo
  {pages} {1694} (\bibinfo {year} {1989})}\BibitemShut {NoStop}%
\bibitem [{\citenamefont {Read}\ and\ \citenamefont {Sachdev}(1990)}]{NRSS90}%
  \BibitemOpen
  \bibfield  {author} {\bibinfo {author} {\bibfnamefont {N.}~\bibnamefont
  {Read}}\ and\ \bibinfo {author} {\bibfnamefont {S.}~\bibnamefont {Sachdev}},\
  }\bibfield  {title} {\enquote {\bibinfo {title} {{Spin-Peierls, valence-bond
  solid, and N\'eel ground states of low-dimensional quantum
  antiferromagnets}},}\ }\href {\doibase 10.1103/PhysRevB.42.4568} {\bibfield
  {journal} {\bibinfo  {journal} {Phys. Rev. B}\ }\textbf {\bibinfo {volume}
  {42}},\ \bibinfo {pages} {4568} (\bibinfo {year} {1990})}\BibitemShut
  {NoStop}%
\bibitem [{\citenamefont {{Sandvik}}(2007)}]{Sandvik07}%
  \BibitemOpen
  \bibfield  {author} {\bibinfo {author} {\bibfnamefont {A.~W.}\ \bibnamefont
  {{Sandvik}}},\ }\bibfield  {title} {\enquote {\bibinfo {title} {{Evidence for
  Deconfined Quantum Criticality in a Two-Dimensional Heisenberg Model with
  Four-Spin Interactions}},}\ }\href {\doibase 10.1103/PhysRevLett.98.227202}
  {\bibfield  {journal} {\bibinfo  {journal} {Phys. Rev. Lett.}\ }\textbf
  {\bibinfo {volume} {98}},\ \bibinfo {eid} {227202} (\bibinfo {year}
  {2007})},\ \Eprint
  {http://arxiv.org/abs/cond-mat/0611343}{cond-mat/0611343}\BibitemShut
  {NoStop}%
\bibitem [{\citenamefont {{Senthil}}\ \emph
  {et~al.}(2004{\natexlab{a}})\citenamefont {{Senthil}}, \citenamefont
  {{Vishwanath}}, \citenamefont {{Balents}}, \citenamefont {{Sachdev}},\ and\
  \citenamefont {{Fisher}}}]{senthil1}%
  \BibitemOpen
  \bibfield  {author} {\bibinfo {author} {\bibfnamefont {T.}~\bibnamefont
  {{Senthil}}}, \bibinfo {author} {\bibfnamefont {A.}~\bibnamefont
  {{Vishwanath}}}, \bibinfo {author} {\bibfnamefont {L.}~\bibnamefont
  {{Balents}}}, \bibinfo {author} {\bibfnamefont {S.}~\bibnamefont
  {{Sachdev}}}, \ and\ \bibinfo {author} {\bibfnamefont {M.~P.~A.}\
  \bibnamefont {{Fisher}}},\ }\bibfield  {title} {\enquote {\bibinfo {title}
  {{Deconfined Quantum Critical Points}},}\ }\href {\doibase
  10.1126/science.1091806} {\bibfield  {journal} {\bibinfo  {journal}
  {Science}\ }\textbf {\bibinfo {volume} {303}},\ \bibinfo {pages} {1490}
  (\bibinfo {year} {2004}{\natexlab{a}})},\ \Eprint
  {http://arxiv.org/abs/cond-mat/0311326}{cond-mat/0311326}\BibitemShut
  {NoStop}%
\bibitem [{\citenamefont {{Vishwanath}}\ \emph {et~al.}(2004)\citenamefont
  {{Vishwanath}}, \citenamefont {{Balents}},\ and\ \citenamefont
  {{Senthil}}}]{AVLBTS04}%
  \BibitemOpen
  \bibfield  {author} {\bibinfo {author} {\bibfnamefont {A.}~\bibnamefont
  {{Vishwanath}}}, \bibinfo {author} {\bibfnamefont {L.}~\bibnamefont
  {{Balents}}}, \ and\ \bibinfo {author} {\bibfnamefont {T.}~\bibnamefont
  {{Senthil}}},\ }\bibfield  {title} {\enquote {\bibinfo {title} {{Quantum
  criticality and deconfinement in phase transitions between valence bond
  solids}},}\ }\href {\doibase 10.1103/PhysRevB.69.224416} {\bibfield
  {journal} {\bibinfo  {journal} {Phys. Rev. B}\ }\textbf {\bibinfo {volume}
  {69}},\ \bibinfo {eid} {224416} (\bibinfo {year} {2004})},\ \Eprint
  {http://arxiv.org/abs/cond-mat/0311085}{cond-mat/0311085}\BibitemShut
  {NoStop}%
\bibitem [{\citenamefont {{Fradkin}}\ \emph {et~al.}(2004)\citenamefont
  {{Fradkin}}, \citenamefont {{Huse}}, \citenamefont {{Moessner}},
  \citenamefont {{Oganesyan}},\ and\ \citenamefont {{Sondhi}}}]{FHMOS04}%
  \BibitemOpen
  \bibfield  {author} {\bibinfo {author} {\bibfnamefont {E.}~\bibnamefont
  {{Fradkin}}}, \bibinfo {author} {\bibfnamefont {D.~A.}\ \bibnamefont
  {{Huse}}}, \bibinfo {author} {\bibfnamefont {R.}~\bibnamefont {{Moessner}}},
  \bibinfo {author} {\bibfnamefont {V.}~\bibnamefont {{Oganesyan}}}, \ and\
  \bibinfo {author} {\bibfnamefont {S.~L.}\ \bibnamefont {{Sondhi}}},\
  }\bibfield  {title} {\enquote {\bibinfo {title} {{Bipartite Rokhsar Kivelson
  points and Cantor deconfinement}},}\ }\href {\doibase
  10.1103/PhysRevB.69.224415} {\bibfield  {journal} {\bibinfo  {journal} {Phys.
  Rev. B}\ }\textbf {\bibinfo {volume} {69}},\ \bibinfo {eid} {224415}
  (\bibinfo {year} {2004})},\ \Eprint
  {http://arxiv.org/abs/cond-mat/0311353}{cond-mat/0311353}\BibitemShut
  {NoStop}%
\bibitem [{\citenamefont {{Hermele}}\ \emph {et~al.}(2004)\citenamefont
  {{Hermele}}, \citenamefont {{Senthil}}, \citenamefont {{Fisher}},
  \citenamefont {{Lee}}, \citenamefont {{Nagaosa}},\ and\ \citenamefont
  {{Wen}}}]{Hermele04}%
  \BibitemOpen
  \bibfield  {author} {\bibinfo {author} {\bibfnamefont {M.}~\bibnamefont
  {{Hermele}}}, \bibinfo {author} {\bibfnamefont {T.}~\bibnamefont
  {{Senthil}}}, \bibinfo {author} {\bibfnamefont {M.~P.~A.}\ \bibnamefont
  {{Fisher}}}, \bibinfo {author} {\bibfnamefont {P.~A.}\ \bibnamefont {{Lee}}},
  \bibinfo {author} {\bibfnamefont {N.}~\bibnamefont {{Nagaosa}}}, \ and\
  \bibinfo {author} {\bibfnamefont {X.-G.}\ \bibnamefont {{Wen}}},\ }\bibfield
  {title} {\enquote {\bibinfo {title} {{Stability of U(1) spin liquids in two
  dimensions}},}\ }\href {\doibase 10.1103/PhysRevB.70.214437} {\bibfield
  {journal} {\bibinfo  {journal} {Phys. Rev. B}\ }\textbf {\bibinfo {volume}
  {70}},\ \bibinfo {eid} {214437} (\bibinfo {year} {2004})},\ \Eprint
  {http://arxiv.org/abs/cond-mat/0404751}{cond-mat/0404751}\BibitemShut
  {NoStop}%
\bibitem [{\citenamefont {Read}\ and\ \citenamefont {Sachdev}(1991)}]{NRSS91}%
  \BibitemOpen
  \bibfield  {author} {\bibinfo {author} {\bibfnamefont {N.}~\bibnamefont
  {Read}}\ and\ \bibinfo {author} {\bibfnamefont {S.}~\bibnamefont {Sachdev}},\
  }\bibfield  {title} {\enquote {\bibinfo {title} {{Large $N$ expansion for
  frustrated quantum antiferromagnets}},}\ }\href {\doibase
  10.1103/PhysRevLett.66.1773} {\bibfield  {journal} {\bibinfo  {journal}
  {Phys. Rev. Lett.}\ }\textbf {\bibinfo {volume} {66}},\ \bibinfo {pages}
  {1773} (\bibinfo {year} {1991})}\BibitemShut {NoStop}%
\bibitem [{\citenamefont {Jalabert}\ and\ \citenamefont
  {Sachdev}(1991)}]{RJSS91}%
  \BibitemOpen
  \bibfield  {author} {\bibinfo {author} {\bibfnamefont {R.~A.}\ \bibnamefont
  {Jalabert}}\ and\ \bibinfo {author} {\bibfnamefont {S.}~\bibnamefont
  {Sachdev}},\ }\bibfield  {title} {\enquote {\bibinfo {title} {{Spontaneous
  alignment of frustrated bonds in an anisotropic, three-dimensional Ising
  model}},}\ }\href {\doibase 10.1103/PhysRevB.44.686} {\bibfield  {journal}
  {\bibinfo  {journal} {Phys. Rev. B}\ }\textbf {\bibinfo {volume} {44}},\
  \bibinfo {pages} {686} (\bibinfo {year} {1991})}\BibitemShut {NoStop}%
\bibitem [{\citenamefont {Wen}(1991)}]{XGW91}%
  \BibitemOpen
  \bibfield  {author} {\bibinfo {author} {\bibfnamefont {X.~G.}\ \bibnamefont
  {Wen}},\ }\bibfield  {title} {\enquote {\bibinfo {title} {Mean-field theory
  of spin-liquid states with finite energy gap and topological orders},}\
  }\href {\doibase 10.1103/PhysRevB.44.2664} {\bibfield  {journal} {\bibinfo
  {journal} {Phys. Rev. B}\ }\textbf {\bibinfo {volume} {44}},\ \bibinfo
  {pages} {2664} (\bibinfo {year} {1991})}\BibitemShut {NoStop}%
\bibitem [{\citenamefont {Sachdev}(1992)}]{SSkagome}%
  \BibitemOpen
  \bibfield  {author} {\bibinfo {author} {\bibfnamefont {S.}~\bibnamefont
  {Sachdev}},\ }\bibfield  {title} {\enquote {\bibinfo {title} {Kagome and
  triangular-lattice heisenberg antiferromagnets: Ordering from quantum
  fluctuations and quantum-disordered ground states with unconfined bosonic
  spinons},}\ }\href {\doibase 10.1103/PhysRevB.45.12377} {\bibfield  {journal}
  {\bibinfo  {journal} {Phys. Rev. B}\ }\textbf {\bibinfo {volume} {45}},\
  \bibinfo {pages} {12377} (\bibinfo {year} {1992})}\BibitemShut {NoStop}%
\bibitem [{\citenamefont {{Sachdev}}\ and\ \citenamefont
  {{Vojta}}(1999)}]{MVSS99}%
  \BibitemOpen
  \bibfield  {author} {\bibinfo {author} {\bibfnamefont {S.}~\bibnamefont
  {{Sachdev}}}\ and\ \bibinfo {author} {\bibfnamefont {M.}~\bibnamefont
  {{Vojta}}},\ }\bibfield  {title} {\enquote {\bibinfo {title} {{Translational
  symmetry breaking in two-dimensional antiferromagnets and
  superconductors}},}\ }\href@noop {} {\bibfield  {journal} {\bibinfo
  {journal} {J. Phys. Soc. Jpn {\bf 69}, Supp. B, 1}\ } (\bibinfo {year}
  {1999})},\ \Eprint
  {http://arxiv.org/abs/cond-mat/9910231}{cond-mat/9910231}\BibitemShut
  {NoStop}%
\bibitem [{\citenamefont {Fradkin}\ and\ \citenamefont
  {Shenker}(1979)}]{EFSS79}%
  \BibitemOpen
  \bibfield  {author} {\bibinfo {author} {\bibfnamefont {E.}~\bibnamefont
  {Fradkin}}\ and\ \bibinfo {author} {\bibfnamefont {S.~H.}\ \bibnamefont
  {Shenker}},\ }\bibfield  {title} {\enquote {\bibinfo {title} {{Phase diagrams
  of lattice gauge theories with Higgs fields}},}\ }\href {\doibase
  10.1103/PhysRevD.19.3682} {\bibfield  {journal} {\bibinfo  {journal} {Phys.
  Rev. D}\ }\textbf {\bibinfo {volume} {19}},\ \bibinfo {pages} {3682}
  (\bibinfo {year} {1979})}\BibitemShut {NoStop}%
\bibitem [{\citenamefont {{Alexander Bais}}\ \emph {et~al.}(1992)\citenamefont
  {{Alexander Bais}}, \citenamefont {{van Driel}},\ and\ \citenamefont {{de
  Wild Propitius}}}]{Bais92}%
  \BibitemOpen
  \bibfield  {author} {\bibinfo {author} {\bibfnamefont {F.}~\bibnamefont
  {{Alexander Bais}}}, \bibinfo {author} {\bibfnamefont {P.}~\bibnamefont {{van
  Driel}}}, \ and\ \bibinfo {author} {\bibfnamefont {M.}~\bibnamefont {{de Wild
  Propitius}}},\ }\bibfield  {title} {\enquote {\bibinfo {title} {{Quantum
  symmetries in discrete gauge theories}},}\ }\href {\doibase
  10.1016/0370-2693(92)90773-W} {\bibfield  {journal} {\bibinfo  {journal}
  {Phys. Lett. B}\ }\textbf {\bibinfo {volume} {280}},\ \bibinfo {pages} {63}
  (\bibinfo {year} {1992})},\ \Eprint
  {http://arxiv.org/abs/hep-th/9203046}{hep-th/9203046}\BibitemShut {NoStop}%
\bibitem [{\citenamefont {{Senthil}}\ and\ \citenamefont
  {{Fisher}}(2000)}]{TSMPAF00}%
  \BibitemOpen
  \bibfield  {author} {\bibinfo {author} {\bibfnamefont {T.}~\bibnamefont
  {{Senthil}}}\ and\ \bibinfo {author} {\bibfnamefont {M.~P.~A.}\ \bibnamefont
  {{Fisher}}},\ }\bibfield  {title} {\enquote {\bibinfo {title} {{Z$_{2}$ gauge
  theory of electron fractionalization in strongly correlated systems}},}\
  }\href {\doibase 10.1103/PhysRevB.62.7850} {\bibfield  {journal} {\bibinfo
  {journal} {Phys. Rev. B}\ }\textbf {\bibinfo {volume} {62}},\ \bibinfo
  {pages} {7850} (\bibinfo {year} {2000})},\ \Eprint
  {http://arxiv.org/abs/cond-mat/9910224}{cond-mat/9910224}\BibitemShut
  {NoStop}%
\bibitem [{\citenamefont {{Moessner}}\ and\ \citenamefont
  {{Sondhi}}(2001)}]{RMSLS01}%
  \BibitemOpen
  \bibfield  {author} {\bibinfo {author} {\bibfnamefont {R.}~\bibnamefont
  {{Moessner}}}\ and\ \bibinfo {author} {\bibfnamefont {S.~L.}\ \bibnamefont
  {{Sondhi}}},\ }\bibfield  {title} {\enquote {\bibinfo {title} {{Resonating
  Valence Bond Phase in the Triangular Lattice Quantum Dimer Model}},}\ }\href
  {\doibase 10.1103/PhysRevLett.86.1881} {\bibfield  {journal} {\bibinfo
  {journal} {Phys. Rev. Lett.}\ }\textbf {\bibinfo {volume} {86}},\ \bibinfo
  {pages} {1881} (\bibinfo {year} {2001})},\ \Eprint
  {http://arxiv.org/abs/cond-mat/0007378}{cond-mat/0007378}\BibitemShut
  {NoStop}%
\bibitem [{\citenamefont {{Kitaev}}(2003)}]{Kitaev03}%
  \BibitemOpen
  \bibfield  {author} {\bibinfo {author} {\bibfnamefont {A.~Y.}\ \bibnamefont
  {{Kitaev}}},\ }\bibfield  {title} {\enquote {\bibinfo {title}
  {{Fault-tolerant quantum computation by anyons}},}\ }\href {\doibase
  10.1016/S0003-4916(02)00018-0} {\bibfield  {journal} {\bibinfo  {journal}
  {Annals of Physics}\ }\textbf {\bibinfo {volume} {303}},\ \bibinfo {pages}
  {2} (\bibinfo {year} {2003})},\ \Eprint
  {http://arxiv.org/abs/quant-ph/9707021}{quant-ph/9707021}\BibitemShut
  {NoStop}%
\bibitem [{\citenamefont {{Wen}}(2003)}]{Wen03}%
  \BibitemOpen
  \bibfield  {author} {\bibinfo {author} {\bibfnamefont {X.-G.}\ \bibnamefont
  {{Wen}}},\ }\bibfield  {title} {\enquote {\bibinfo {title} {{Quantum Orders
  in an Exact Soluble Model}},}\ }\href {\doibase
  10.1103/PhysRevLett.90.016803} {\bibfield  {journal} {\bibinfo  {journal}
  {Phys. Rev. Lett.}\ }\textbf {\bibinfo {volume} {90}},\ \bibinfo {eid}
  {016803} (\bibinfo {year} {2003})},\ \Eprint
  {http://arxiv.org/abs/quant-ph/0205004}{quant-ph/0205004}\BibitemShut
  {NoStop}%
\bibitem [{\citenamefont {{Freedman}}\ \emph {et~al.}(2004)\citenamefont
  {{Freedman}}, \citenamefont {{Nayak}}, \citenamefont {{Shtengel}},
  \citenamefont {{Walker}},\ and\ \citenamefont {{Wang}}}]{Freedman04}%
  \BibitemOpen
  \bibfield  {author} {\bibinfo {author} {\bibfnamefont {M.}~\bibnamefont
  {{Freedman}}}, \bibinfo {author} {\bibfnamefont {C.}~\bibnamefont {{Nayak}}},
  \bibinfo {author} {\bibfnamefont {K.}~\bibnamefont {{Shtengel}}}, \bibinfo
  {author} {\bibfnamefont {K.}~\bibnamefont {{Walker}}}, \ and\ \bibinfo
  {author} {\bibfnamefont {Z.}~\bibnamefont {{Wang}}},\ }\bibfield  {title}
  {\enquote {\bibinfo {title} {{A class of P, T-invariant topological phases of
  interacting electrons}},}\ }\href {\doibase 10.1016/j.aop.2004.01.006}
  {\bibfield  {journal} {\bibinfo  {journal} {Annals of Physics}\ }\textbf
  {\bibinfo {volume} {310}},\ \bibinfo {pages} {428} (\bibinfo {year}
  {2004})},\ \Eprint
  {http://arxiv.org/abs/cond-mat/0307511}{cond-mat/0307511}\BibitemShut
  {NoStop}%
\bibitem [{\citenamefont {{Kane}}\ and\ \citenamefont
  {{Mele}}(2005)}]{KaneMele05}%
  \BibitemOpen
  \bibfield  {author} {\bibinfo {author} {\bibfnamefont {C.~L.}\ \bibnamefont
  {{Kane}}}\ and\ \bibinfo {author} {\bibfnamefont {E.~J.}\ \bibnamefont
  {{Mele}}},\ }\bibfield  {title} {\enquote {\bibinfo {title}
  {{$\mathbb{Z}_{2}$ Topological Order and the Quantum Spin Hall Effect}},}\
  }\href {\doibase 10.1103/PhysRevLett.95.146802} {\bibfield  {journal}
  {\bibinfo  {journal} {Physical Review Letters}\ }\textbf {\bibinfo {volume}
  {95}},\ \bibinfo {eid} {146802} (\bibinfo {year} {2005})},\ \Eprint
  {http://arxiv.org/abs/cond-mat/0506581}{cond-mat/0506581}\BibitemShut
  {NoStop}%
\bibitem [{\citenamefont {{Lu}}\ and\ \citenamefont
  {{Vishwanath}}(2012)}]{LuVishwanath12}%
  \BibitemOpen
  \bibfield  {author} {\bibinfo {author} {\bibfnamefont {Y.-M.}\ \bibnamefont
  {{Lu}}}\ and\ \bibinfo {author} {\bibfnamefont {A.}~\bibnamefont
  {{Vishwanath}}},\ }\bibfield  {title} {\enquote {\bibinfo {title} {{Theory
  and classification of interacting integer topological phases in two
  dimensions: A Chern-Simons approach}},}\ }\href {\doibase
  10.1103/PhysRevB.86.125119} {\bibfield  {journal} {\bibinfo  {journal} {Phys.
  Rev. B}\ }\textbf {\bibinfo {volume} {86}},\ \bibinfo {eid} {125119}
  (\bibinfo {year} {2012})},\ \Eprint
  {http://arxiv.org/abs/1205.3156}{arXiv:1205.3156
  [cond-mat.str-el]}\BibitemShut {NoStop}%
\bibitem [{\citenamefont {{Plat{\'e}}}\ \emph {et~al.}(2005)\citenamefont
  {{Plat{\'e}}}, \citenamefont {{Mottershead}}, \citenamefont {{Elfimov}},
  \citenamefont {{Peets}}, \citenamefont {{Liang}}, \citenamefont {{Bonn}},
  \citenamefont {{Hardy}}, \citenamefont {{Chiuzbaian}}, \citenamefont
  {{Falub}}, \citenamefont {{Shi}}, \citenamefont {{Patthey}},\ and\
  \citenamefont {{Damascelli}}}]{Dama05}%
  \BibitemOpen
  \bibfield  {author} {\bibinfo {author} {\bibfnamefont {M.}~\bibnamefont
  {{Plat{\'e}}}}, \bibinfo {author} {\bibfnamefont {J.~D.}\ \bibnamefont
  {{Mottershead}}}, \bibinfo {author} {\bibfnamefont {I.~S.}\ \bibnamefont
  {{Elfimov}}}, \bibinfo {author} {\bibfnamefont {D.~C.}\ \bibnamefont
  {{Peets}}}, \bibinfo {author} {\bibfnamefont {R.}~\bibnamefont {{Liang}}},
  \bibinfo {author} {\bibfnamefont {D.~A.}\ \bibnamefont {{Bonn}}}, \bibinfo
  {author} {\bibfnamefont {W.~N.}\ \bibnamefont {{Hardy}}}, \bibinfo {author}
  {\bibfnamefont {S.}~\bibnamefont {{Chiuzbaian}}}, \bibinfo {author}
  {\bibfnamefont {M.}~\bibnamefont {{Falub}}}, \bibinfo {author} {\bibfnamefont
  {M.}~\bibnamefont {{Shi}}}, \bibinfo {author} {\bibfnamefont
  {L.}~\bibnamefont {{Patthey}}}, \ and\ \bibinfo {author} {\bibfnamefont
  {A.}~\bibnamefont {{Damascelli}}},\ }\bibfield  {title} {\enquote {\bibinfo
  {title} {{Fermi Surface and Quasiparticle Excitations of Overdoped
  Tl$_{2}$Ba$_{2}$CuO$_{6+\delta}$}},}\ }\href {\doibase
  10.1103/PhysRevLett.95.077001} {\bibfield  {journal} {\bibinfo  {journal}
  {Phys. Rev. Lett.}\ }\textbf {\bibinfo {volume} {95}},\ \bibinfo {eid}
  {077001} (\bibinfo {year} {2005})},\ \Eprint
  {http://arxiv.org/abs/cond-mat/0503117}{cond-mat/0503117}\BibitemShut
  {NoStop}%
\bibitem [{\citenamefont {{Senthil}}\ \emph
  {et~al.}(2004{\natexlab{b}})\citenamefont {{Senthil}}, \citenamefont
  {{Vojta}},\ and\ \citenamefont {{Sachdev}}}]{TSMVSS04}%
  \BibitemOpen
  \bibfield  {author} {\bibinfo {author} {\bibfnamefont {T.}~\bibnamefont
  {{Senthil}}}, \bibinfo {author} {\bibfnamefont {M.}~\bibnamefont {{Vojta}}},
  \ and\ \bibinfo {author} {\bibfnamefont {S.}~\bibnamefont {{Sachdev}}},\
  }\bibfield  {title} {\enquote {\bibinfo {title} {{Weak magnetism and
  non-Fermi liquids near heavy-fermion critical points}},}\ }\href {\doibase
  10.1103/PhysRevB.69.035111} {\bibfield  {journal} {\bibinfo  {journal} {Phys.
  Rev. B}\ }\textbf {\bibinfo {volume} {69}},\ \bibinfo {eid} {035111}
  (\bibinfo {year} {2004}{\natexlab{b}})},\ \Eprint
  {http://arxiv.org/abs/cond-mat/0305193}{cond-mat/0305193}\BibitemShut
  {NoStop}%
\bibitem [{\citenamefont {{Paramekanti}}\ and\ \citenamefont
  {{Vishwanath}}(2004)}]{APAV04}%
  \BibitemOpen
  \bibfield  {author} {\bibinfo {author} {\bibfnamefont {A.}~\bibnamefont
  {{Paramekanti}}}\ and\ \bibinfo {author} {\bibfnamefont {A.}~\bibnamefont
  {{Vishwanath}}},\ }\bibfield  {title} {\enquote {\bibinfo {title} {{Extending
  Luttinger's theorem to $\mathbb{Z}_{2}$ fractionalized phases of matter}},}\
  }\href {\doibase 10.1103/PhysRevB.70.245118} {\bibfield  {journal} {\bibinfo
  {journal} {Phys. Rev. B}\ }\textbf {\bibinfo {volume} {70}},\ \bibinfo {eid}
  {245118} (\bibinfo {year} {2004})},\ \Eprint
  {http://arxiv.org/abs/cond-mat/0406619}{cond-mat/0406619}\BibitemShut
  {NoStop}%
\bibitem [{\citenamefont {{Yang}}\ \emph {et~al.}(2006)\citenamefont {{Yang}},
  \citenamefont {{Rice}},\ and\ \citenamefont {{Zhang}}}]{YRZ}%
  \BibitemOpen
  \bibfield  {author} {\bibinfo {author} {\bibfnamefont {K.-Y.}\ \bibnamefont
  {{Yang}}}, \bibinfo {author} {\bibfnamefont {T.~M.}\ \bibnamefont {{Rice}}},
  \ and\ \bibinfo {author} {\bibfnamefont {F.-C.}\ \bibnamefont {{Zhang}}},\
  }\bibfield  {title} {\enquote {\bibinfo {title} {{Phenomenological theory of
  the pseudogap state}},}\ }\href {\doibase 10.1103/PhysRevB.73.174501}
  {\bibfield  {journal} {\bibinfo  {journal} {Phys. Rev. B}\ }\textbf {\bibinfo
  {volume} {73}},\ \bibinfo {eid} {174501} (\bibinfo {year} {2006})},\ \Eprint
  {http://arxiv.org/abs/cond-mat/0602164}{cond-mat/0602164}\BibitemShut
  {NoStop}%
\bibitem [{\citenamefont {Kivelson}\ \emph {et~al.}(1987)\citenamefont
  {Kivelson}, \citenamefont {Rokhsar},\ and\ \citenamefont {Sethna}}]{KRS87}%
  \BibitemOpen
  \bibfield  {author} {\bibinfo {author} {\bibfnamefont {S.~A.}\ \bibnamefont
  {Kivelson}}, \bibinfo {author} {\bibfnamefont {D.~S.}\ \bibnamefont
  {Rokhsar}}, \ and\ \bibinfo {author} {\bibfnamefont {J.~P.}\ \bibnamefont
  {Sethna}},\ }\bibfield  {title} {\enquote {\bibinfo {title} {{Topology of the
  resonating valence-bond state: Solitons and high-${T}_{c}$
  superconductivity}},}\ }\href {\doibase 10.1103/PhysRevB.35.8865} {\bibfield
  {journal} {\bibinfo  {journal} {Phys. Rev. B}\ }\textbf {\bibinfo {volume}
  {35}},\ \bibinfo {pages} {8865} (\bibinfo {year} {1987})}\BibitemShut
  {NoStop}%
\bibitem [{\citenamefont {Read}\ and\ \citenamefont
  {Chakraborty}(1989)}]{RC89}%
  \BibitemOpen
  \bibfield  {author} {\bibinfo {author} {\bibfnamefont {N.}~\bibnamefont
  {Read}}\ and\ \bibinfo {author} {\bibfnamefont {B.}~\bibnamefont
  {Chakraborty}},\ }\bibfield  {title} {\enquote {\bibinfo {title} {Statistics
  of the excitations of the resonating-valence-bond state},}\ }\href {\doibase
  10.1103/PhysRevB.40.7133} {\bibfield  {journal} {\bibinfo  {journal} {Phys.
  Rev. B}\ }\textbf {\bibinfo {volume} {40}},\ \bibinfo {pages} {7133}
  (\bibinfo {year} {1989})}\BibitemShut {NoStop}%
\bibitem [{\citenamefont {Punk}\ \emph {et~al.}(2015)\citenamefont {Punk},
  \citenamefont {Allais},\ and\ \citenamefont {Sachdev}}]{Punk15}%
  \BibitemOpen
  \bibfield  {author} {\bibinfo {author} {\bibfnamefont {M.}~\bibnamefont
  {Punk}}, \bibinfo {author} {\bibfnamefont {A.}~\bibnamefont {Allais}}, \ and\
  \bibinfo {author} {\bibfnamefont {S.}~\bibnamefont {Sachdev}},\ }\bibfield
  {title} {\enquote {\bibinfo {title} {{A quantum dimer model for the pseudogap
  metal}},}\ }\href {\doibase 10.1073/pnas.1512206112} {\bibfield  {journal}
  {\bibinfo  {journal} {Proc. Nat. Acad. Sci.}\ }\textbf {\bibinfo {volume}
  {112}},\ \bibinfo {pages} {9552} (\bibinfo {year} {2015})},\ \Eprint
  {http://arxiv.org/abs/1501.00978}{arXiv:1501.00978
  [cond-mat.str-el]}\BibitemShut {NoStop}%
%%CITATION = ARXIV:1501.00978;%%
\bibitem [{\citenamefont {{Ferrero}}\ \emph {et~al.}(2009)\citenamefont
  {{Ferrero}}, \citenamefont {{Cornaglia}}, \citenamefont {{de Leo}},
  \citenamefont {{Parcollet}}, \citenamefont {{Kotliar}},\ and\ \citenamefont
  {{Georges}}}]{PG09}%
  \BibitemOpen
  \bibfield  {author} {\bibinfo {author} {\bibfnamefont {M.}~\bibnamefont
  {{Ferrero}}}, \bibinfo {author} {\bibfnamefont {P.~S.}\ \bibnamefont
  {{Cornaglia}}}, \bibinfo {author} {\bibfnamefont {L.}~\bibnamefont {{de
  Leo}}}, \bibinfo {author} {\bibfnamefont {O.}~\bibnamefont {{Parcollet}}},
  \bibinfo {author} {\bibfnamefont {G.}~\bibnamefont {{Kotliar}}}, \ and\
  \bibinfo {author} {\bibfnamefont {A.}~\bibnamefont {{Georges}}},\ }\bibfield
  {title} {\enquote {\bibinfo {title} {{Pseudogap opening and formation of
  Fermi arcs as an orbital-selective Mott transition in momentum space}},}\
  }\href {\doibase 10.1103/PhysRevB.80.064501} {\bibfield  {journal} {\bibinfo
  {journal} {Phys. Rev. B}\ }\textbf {\bibinfo {volume} {80}},\ \bibinfo {eid}
  {064501} (\bibinfo {year} {2009})},\ \Eprint
  {http://arxiv.org/abs/0903.2480}{arXiv:0903.2480
  [cond-mat.str-el]}\BibitemShut {NoStop}%
\bibitem [{\citenamefont {{Sordi}}\ \emph {et~al.}(2011)\citenamefont
  {{Sordi}}, \citenamefont {{Haule}},\ and\ \citenamefont
  {{Tremblay}}}]{AMT11}%
  \BibitemOpen
  \bibfield  {author} {\bibinfo {author} {\bibfnamefont {G.}~\bibnamefont
  {{Sordi}}}, \bibinfo {author} {\bibfnamefont {K.}~\bibnamefont {{Haule}}}, \
  and\ \bibinfo {author} {\bibfnamefont {A.-M.~S.}\ \bibnamefont
  {{Tremblay}}},\ }\bibfield  {title} {\enquote {\bibinfo {title} {{Mott
  physics and first-order transition between two metals in the normal-state
  phase diagram of the two-dimensional Hubbard model}},}\ }\href {\doibase
  10.1103/PhysRevB.84.075161} {\bibfield  {journal} {\bibinfo  {journal} {Phys.
  Rev. B}\ }\textbf {\bibinfo {volume} {84}},\ \bibinfo {eid} {075161}
  (\bibinfo {year} {2011})},\ \Eprint
  {http://arxiv.org/abs/1102.0463}{arXiv:1102.0463
  [cond-mat.str-el]}\BibitemShut {NoStop}%
\bibitem [{\citenamefont {{Senthil}}\ \emph {et~al.}(2003)\citenamefont
  {{Senthil}}, \citenamefont {{Sachdev}},\ and\ \citenamefont
  {{Vojta}}}]{TSSSMV03}%
  \BibitemOpen
  \bibfield  {author} {\bibinfo {author} {\bibfnamefont {T.}~\bibnamefont
  {{Senthil}}}, \bibinfo {author} {\bibfnamefont {S.}~\bibnamefont
  {{Sachdev}}}, \ and\ \bibinfo {author} {\bibfnamefont {M.}~\bibnamefont
  {{Vojta}}},\ }\bibfield  {title} {\enquote {\bibinfo {title} {{Fractionalized
  Fermi Liquids}},}\ }\href {\doibase 10.1103/PhysRevLett.90.216403} {\bibfield
   {journal} {\bibinfo  {journal} {Phys. Rev. Lett.}\ }\textbf {\bibinfo
  {volume} {90}},\ \bibinfo {eid} {216403} (\bibinfo {year} {2003})},\ \Eprint
  {http://arxiv.org/abs/cond-mat/0209144}{cond-mat/0209144}\BibitemShut
  {NoStop}%
\bibitem [{\citenamefont {{Punk}}\ and\ \citenamefont
  {{Sachdev}}(2012)}]{MPSS12}%
  \BibitemOpen
  \bibfield  {author} {\bibinfo {author} {\bibfnamefont {M.}~\bibnamefont
  {{Punk}}}\ and\ \bibinfo {author} {\bibfnamefont {S.}~\bibnamefont
  {{Sachdev}}},\ }\bibfield  {title} {\enquote {\bibinfo {title} {{Fermi
  surface reconstruction in hole-doped $t$-$J$ models without long-range
  antiferromagnetic order}},}\ }\href {\doibase 10.1103/PhysRevB.85.195123}
  {\bibfield  {journal} {\bibinfo  {journal} {Phys. Rev. B}\ }\textbf {\bibinfo
  {volume} {85}},\ \bibinfo {eid} {195123} (\bibinfo {year} {2012})},\ \Eprint
  {http://arxiv.org/abs/1202.4023}{arXiv:1202.4023
  [cond-mat.str-el]}\BibitemShut {NoStop}%
\bibitem [{\citenamefont {Chubukov}\ and\ \citenamefont
  {Sachdev}(1993)}]{CS93}%
  \BibitemOpen
  \bibfield  {author} {\bibinfo {author} {\bibfnamefont {A.~V.}\ \bibnamefont
  {Chubukov}}\ and\ \bibinfo {author} {\bibfnamefont {S.}~\bibnamefont
  {Sachdev}},\ }\bibfield  {title} {\enquote {\bibinfo {title} {{Chubukov and
  Sachdev reply}},}\ }\href {\doibase 10.1103/PhysRevLett.71.3615} {\bibfield
  {journal} {\bibinfo  {journal} {Phys. Rev. Lett.}\ }\textbf {\bibinfo
  {volume} {71}},\ \bibinfo {pages} {3615} (\bibinfo {year}
  {1993})}\BibitemShut {NoStop}%
\bibitem [{\citenamefont {{Sachdev}}(1994)}]{SS94}%
  \BibitemOpen
  \bibfield  {author} {\bibinfo {author} {\bibfnamefont {S.}~\bibnamefont
  {{Sachdev}}},\ }\bibfield  {title} {\enquote {\bibinfo {title} {{Quantum
  phases of the Shraiman-Siggia model}},}\ }\href {\doibase
  10.1103/PhysRevB.49.6770} {\bibfield  {journal} {\bibinfo  {journal} {Phys.
  Rev. B}\ }\textbf {\bibinfo {volume} {49}},\ \bibinfo {pages} {6770}
  (\bibinfo {year} {1994})},\ \Eprint
  {http://arxiv.org/abs/cond-mat/9311037}{cond-mat/9311037}\BibitemShut
  {NoStop}%
\bibitem [{\citenamefont {{Wen}}\ and\ \citenamefont {{Lee}}(1996)}]{XGWPAL96}%
  \BibitemOpen
  \bibfield  {author} {\bibinfo {author} {\bibfnamefont {X.-G.}\ \bibnamefont
  {{Wen}}}\ and\ \bibinfo {author} {\bibfnamefont {P.~A.}\ \bibnamefont
  {{Lee}}},\ }\bibfield  {title} {\enquote {\bibinfo {title} {{Theory of
  Underdoped Cuprates}},}\ }\href {\doibase 10.1103/PhysRevLett.76.503}
  {\bibfield  {journal} {\bibinfo  {journal} {Phys. Rev. Lett.}\ }\textbf
  {\bibinfo {volume} {76}},\ \bibinfo {pages} {503} (\bibinfo {year} {1996})},\
  \Eprint {http://arxiv.org/abs/cond-mat/9506065}{cond-mat/9506065}\BibitemShut
  {NoStop}%
\bibitem [{\citenamefont {{Ribeiro}}\ and\ \citenamefont
  {{Wen}}(2006)}]{XGW06}%
  \BibitemOpen
  \bibfield  {author} {\bibinfo {author} {\bibfnamefont {T.~C.}\ \bibnamefont
  {{Ribeiro}}}\ and\ \bibinfo {author} {\bibfnamefont {X.-G.}\ \bibnamefont
  {{Wen}}},\ }\bibfield  {title} {\enquote {\bibinfo {title} {{Doped carrier
  formulation and mean-field theory of the $tt{'}t{''}J$ model}},}\ }\href
  {\doibase 10.1103/PhysRevB.74.155113} {\bibfield  {journal} {\bibinfo
  {journal} {Phys. Rev. B}\ }\textbf {\bibinfo {volume} {74}},\ \bibinfo {eid}
  {155113} (\bibinfo {year} {2006})},\ \Eprint
  {http://arxiv.org/abs/cond-mat/0601174}{cond-mat/0601174}\BibitemShut
  {NoStop}%
\bibitem [{\citenamefont {{Kaul}}\ \emph {et~al.}(2007)\citenamefont {{Kaul}},
  \citenamefont {{Kolezhuk}}, \citenamefont {{Levin}}, \citenamefont
  {{Sachdev}},\ and\ \citenamefont {{Senthil}}}]{RKK07}%
  \BibitemOpen
  \bibfield  {author} {\bibinfo {author} {\bibfnamefont {R.~K.}\ \bibnamefont
  {{Kaul}}}, \bibinfo {author} {\bibfnamefont {A.}~\bibnamefont {{Kolezhuk}}},
  \bibinfo {author} {\bibfnamefont {M.}~\bibnamefont {{Levin}}}, \bibinfo
  {author} {\bibfnamefont {S.}~\bibnamefont {{Sachdev}}}, \ and\ \bibinfo
  {author} {\bibfnamefont {T.}~\bibnamefont {{Senthil}}},\ }\bibfield  {title}
  {\enquote {\bibinfo {title} {{Hole dynamics in an antiferromagnet across a
  deconfined quantum critical point}},}\ }\href {\doibase
  10.1103/PhysRevB.75.235122} {\bibfield  {journal} {\bibinfo  {journal} {Phys.
  Rev. B}\ }\textbf {\bibinfo {volume} {75}},\ \bibinfo {eid} {235122}
  (\bibinfo {year} {2007})},\ \Eprint
  {http://arxiv.org/abs/cond-mat/0702119}{cond-mat/0702119}\BibitemShut
  {NoStop}%
\bibitem [{\citenamefont {{Kaul}}\ \emph {et~al.}(2008)\citenamefont {{Kaul}},
  \citenamefont {{Kim}}, \citenamefont {{Sachdev}},\ and\ \citenamefont
  {{Senthil}}}]{RKK08}%
  \BibitemOpen
  \bibfield  {author} {\bibinfo {author} {\bibfnamefont {R.~K.}\ \bibnamefont
  {{Kaul}}}, \bibinfo {author} {\bibfnamefont {Y.~B.}\ \bibnamefont {{Kim}}},
  \bibinfo {author} {\bibfnamefont {S.}~\bibnamefont {{Sachdev}}}, \ and\
  \bibinfo {author} {\bibfnamefont {T.}~\bibnamefont {{Senthil}}},\ }\bibfield
  {title} {\enquote {\bibinfo {title} {{Algebraic charge liquids}},}\ }\href
  {\doibase 10.1038/nphys790} {\bibfield  {journal} {\bibinfo  {journal}
  {Nature Physics}\ }\textbf {\bibinfo {volume} {4}},\ \bibinfo {pages} {28}
  (\bibinfo {year} {2008})},\ \Eprint
  {http://arxiv.org/abs/0706.2187}{arXiv:0706.2187
  [cond-mat.str-el]}\BibitemShut {NoStop}%
\bibitem [{\citenamefont {{Qi}}\ and\ \citenamefont
  {{Sachdev}}(2010)}]{YQSS10}%
  \BibitemOpen
  \bibfield  {author} {\bibinfo {author} {\bibfnamefont {Y.}~\bibnamefont
  {{Qi}}}\ and\ \bibinfo {author} {\bibfnamefont {S.}~\bibnamefont
  {{Sachdev}}},\ }\bibfield  {title} {\enquote {\bibinfo {title} {{Effective
  theory of Fermi pockets in fluctuating antiferromagnets}},}\ }\href {\doibase
  10.1103/PhysRevB.81.115129} {\bibfield  {journal} {\bibinfo  {journal} {Phys.
  Rev. B}\ }\textbf {\bibinfo {volume} {81}},\ \bibinfo {eid} {115129}
  (\bibinfo {year} {2010})},\ \Eprint
  {http://arxiv.org/abs/0912.0943}{arXiv:0912.0943
  [cond-mat.str-el]}\BibitemShut {NoStop}%
\bibitem [{\citenamefont {{Mei}}\ \emph {et~al.}(2012)\citenamefont {{Mei}},
  \citenamefont {{Kawasaki}}, \citenamefont {{Zheng}}, \citenamefont {{Weng}},\
  and\ \citenamefont {{Wen}}}]{Mei12}%
  \BibitemOpen
  \bibfield  {author} {\bibinfo {author} {\bibfnamefont {J.-W.}\ \bibnamefont
  {{Mei}}}, \bibinfo {author} {\bibfnamefont {S.}~\bibnamefont {{Kawasaki}}},
  \bibinfo {author} {\bibfnamefont {G.-Q.}\ \bibnamefont {{Zheng}}}, \bibinfo
  {author} {\bibfnamefont {Z.-Y.}\ \bibnamefont {{Weng}}}, \ and\ \bibinfo
  {author} {\bibfnamefont {X.-G.}\ \bibnamefont {{Wen}}},\ }\bibfield  {title}
  {\enquote {\bibinfo {title} {{Luttinger-volume violating Fermi liquid in the
  pseudogap phase of the cuprate superconductors}},}\ }\href {\doibase
  10.1103/PhysRevB.85.134519} {\bibfield  {journal} {\bibinfo  {journal} {Phys.
  Rev. B}\ }\textbf {\bibinfo {volume} {85}},\ \bibinfo {eid} {134519}
  (\bibinfo {year} {2012})},\ \Eprint
  {http://arxiv.org/abs/1109.0406}{arXiv:1109.0406
  [cond-mat.supr-con]}\BibitemShut {NoStop}%
\bibitem [{\citenamefont {{Ferraz}}\ and\ \citenamefont
  {{Kochetov}}(2013)}]{Ferraz13}%
  \BibitemOpen
  \bibfield  {author} {\bibinfo {author} {\bibfnamefont {A.}~\bibnamefont
  {{Ferraz}}}\ and\ \bibinfo {author} {\bibfnamefont {E.}~\bibnamefont
  {{Kochetov}}},\ }\bibfield  {title} {\enquote {\bibinfo {title} {{Gauge
  invariance and spinon-dopon confinement in the t-J model: implications for
  Fermi surface reconstruction in the cuprates}},}\ }\href {\doibase
  10.1140/epjb/e2013-40849-8} {\bibfield  {journal} {\bibinfo  {journal} {Eur.
  Phys. J. B}\ }\textbf {\bibinfo {volume} {86}},\ \bibinfo {eid} {512}
  (\bibinfo {year} {2013})},\ \Eprint
  {http://arxiv.org/abs/1312.6167}{arXiv:1312.6167
  [cond-mat.str-el]}\BibitemShut {NoStop}%
\bibitem [{\citenamefont {{Oshikawa}}(2000)}]{MO00}%
  \BibitemOpen
  \bibfield  {author} {\bibinfo {author} {\bibfnamefont {M.}~\bibnamefont
  {{Oshikawa}}},\ }\bibfield  {title} {\enquote {\bibinfo {title} {{Topological
  Approach to Luttinger's Theorem and the Fermi Surface of a Kondo Lattice}},}\
  }\href {\doibase 10.1103/PhysRevLett.84.3370} {\bibfield  {journal} {\bibinfo
   {journal} {Phys. Rev. Lett.}\ }\textbf {\bibinfo {volume} {84}},\ \bibinfo
  {pages} {3370} (\bibinfo {year} {2000})},\ \Eprint
  {http://arxiv.org/abs/cond-mat/0002392}{cond-mat/0002392}\BibitemShut
  {NoStop}%
\bibitem [{\citenamefont {Shen}\ \emph {et~al.}(2005)\citenamefont {Shen},
  \citenamefont {Ronning}, \citenamefont {Lu}, \citenamefont {Baumberger},
  \citenamefont {Ingle}, \citenamefont {Lee}, \citenamefont {Meevasana},
  \citenamefont {Kohsaka}, \citenamefont {Azuma}, \citenamefont {Takano},
  \citenamefont {Takagi},\ and\ \citenamefont {Shen}}]{Shen05}%
  \BibitemOpen
  \bibfield  {author} {\bibinfo {author} {\bibfnamefont {K.~M.}\ \bibnamefont
  {Shen}}, \bibinfo {author} {\bibfnamefont {F.}~\bibnamefont {Ronning}},
  \bibinfo {author} {\bibfnamefont {D.~H.}\ \bibnamefont {Lu}}, \bibinfo
  {author} {\bibfnamefont {F.}~\bibnamefont {Baumberger}}, \bibinfo {author}
  {\bibfnamefont {N.~J.~C.}\ \bibnamefont {Ingle}}, \bibinfo {author}
  {\bibfnamefont {W.~S.}\ \bibnamefont {Lee}}, \bibinfo {author} {\bibfnamefont
  {W.}~\bibnamefont {Meevasana}}, \bibinfo {author} {\bibfnamefont
  {Y.}~\bibnamefont {Kohsaka}}, \bibinfo {author} {\bibfnamefont
  {M.}~\bibnamefont {Azuma}}, \bibinfo {author} {\bibfnamefont
  {M.}~\bibnamefont {Takano}}, \bibinfo {author} {\bibfnamefont
  {H.}~\bibnamefont {Takagi}}, \ and\ \bibinfo {author} {\bibfnamefont {Z.-X.}\
  \bibnamefont {Shen}},\ }\bibfield  {title} {\enquote {\bibinfo {title}
  {{Nodal Quasiparticles and Antinodal Charge Ordering in
  Ca$_{2-x}$Na$_x$CuO$_2$Cl$_2$}},}\ }\href {\doibase 10.1126/science.1103627}
  {\bibfield  {journal} {\bibinfo  {journal} {Science}\ }\textbf {\bibinfo
  {volume} {307}},\ \bibinfo {pages} {901} (\bibinfo {year}
  {2005})}\BibitemShut {NoStop}%
\bibitem [{\citenamefont {Yang}\ \emph {et~al.}(2011)\citenamefont {Yang},
  \citenamefont {Rameau}, \citenamefont {Pan}, \citenamefont {Gu},
  \citenamefont {Johnson}, \citenamefont {Claus}, \citenamefont {Hinks},\ and\
  \citenamefont {Kidd}}]{Johnson10}%
  \BibitemOpen
  \bibfield  {author} {\bibinfo {author} {\bibfnamefont {H.-B.}\ \bibnamefont
  {Yang}}, \bibinfo {author} {\bibfnamefont {J.~D.}\ \bibnamefont {Rameau}},
  \bibinfo {author} {\bibfnamefont {Z.-H.}\ \bibnamefont {Pan}}, \bibinfo
  {author} {\bibfnamefont {G.~D.}\ \bibnamefont {Gu}}, \bibinfo {author}
  {\bibfnamefont {P.~D.}\ \bibnamefont {Johnson}}, \bibinfo {author}
  {\bibfnamefont {H.}~\bibnamefont {Claus}}, \bibinfo {author} {\bibfnamefont
  {D.~G.}\ \bibnamefont {Hinks}}, \ and\ \bibinfo {author} {\bibfnamefont
  {T.~E.}\ \bibnamefont {Kidd}},\ }\bibfield  {title} {\enquote {\bibinfo
  {title} {{Reconstructed Fermi Surface of Underdoped
  ${\mathrm{Bi}}_{2}{\mathrm{Sr}}_{2}{\mathrm{CaCu}}_{2}{\mathrm{O}}_{8+\ensuremath{\delta}}$
  Cuprate Superconductors}},}\ }\href {\doibase 10.1103/PhysRevLett.107.047003}
  {\bibfield  {journal} {\bibinfo  {journal} {Phys. Rev. Lett.}\ }\textbf
  {\bibinfo {volume} {107}},\ \bibinfo {pages} {047003} (\bibinfo {year}
  {2011})},\ \Eprint {http://arxiv.org/abs/1008.3121}{arXiv:1008.3121
  [cond-mat.supr-con]}\BibitemShut {NoStop}%
\bibitem [{\citenamefont {{Ando}}\ \emph {et~al.}(2004)\citenamefont {{Ando}},
  \citenamefont {{Kurita}}, \citenamefont {{Komiya}}, \citenamefont {{Ono}},\
  and\ \citenamefont {{Segawa}}}]{Ando04}%
  \BibitemOpen
  \bibfield  {author} {\bibinfo {author} {\bibfnamefont {Y.}~\bibnamefont
  {{Ando}}}, \bibinfo {author} {\bibfnamefont {Y.}~\bibnamefont {{Kurita}}},
  \bibinfo {author} {\bibfnamefont {S.}~\bibnamefont {{Komiya}}}, \bibinfo
  {author} {\bibfnamefont {S.}~\bibnamefont {{Ono}}}, \ and\ \bibinfo {author}
  {\bibfnamefont {K.}~\bibnamefont {{Segawa}}},\ }\bibfield  {title} {\enquote
  {\bibinfo {title} {{Evolution of the Hall Coefficient and the Peculiar
  Electronic Structure of the Cuprate Superconductors}},}\ }\href {\doibase
  10.1103/PhysRevLett.92.197001} {\bibfield  {journal} {\bibinfo  {journal}
  {Phys. Rev. Lett.}\ }\textbf {\bibinfo {volume} {92}},\ \bibinfo {eid}
  {197001} (\bibinfo {year} {2004})},\ \Eprint
  {http://arxiv.org/abs/cond-mat/0401034}{cond-mat/0401034}\BibitemShut
  {NoStop}%
\bibitem [{\citenamefont {{Kohsaka}}\ \emph {et~al.}(2007)\citenamefont
  {{Kohsaka}}, \citenamefont {{Taylor}}, \citenamefont {{Fujita}},
  \citenamefont {{Schmidt}}, \citenamefont {{Lupien}}, \citenamefont
  {{Hanaguri}}, \citenamefont {{Azuma}}, \citenamefont {{Takano}},
  \citenamefont {{Eisaki}}, \citenamefont {{Takagi}}, \citenamefont
  {{Uchida}},\ and\ \citenamefont {{S{\'e}amus Davis}}}]{Kohsaka07}%
  \BibitemOpen
  \bibfield  {author} {\bibinfo {author} {\bibfnamefont {Y.}~\bibnamefont
  {{Kohsaka}}}, \bibinfo {author} {\bibfnamefont {C.}~\bibnamefont {{Taylor}}},
  \bibinfo {author} {\bibfnamefont {K.}~\bibnamefont {{Fujita}}}, \bibinfo
  {author} {\bibfnamefont {A.}~\bibnamefont {{Schmidt}}}, \bibinfo {author}
  {\bibfnamefont {C.}~\bibnamefont {{Lupien}}}, \bibinfo {author}
  {\bibfnamefont {T.}~\bibnamefont {{Hanaguri}}}, \bibinfo {author}
  {\bibfnamefont {M.}~\bibnamefont {{Azuma}}}, \bibinfo {author} {\bibfnamefont
  {M.}~\bibnamefont {{Takano}}}, \bibinfo {author} {\bibfnamefont
  {H.}~\bibnamefont {{Eisaki}}}, \bibinfo {author} {\bibfnamefont
  {H.}~\bibnamefont {{Takagi}}}, \bibinfo {author} {\bibfnamefont
  {S.}~\bibnamefont {{Uchida}}}, \ and\ \bibinfo {author} {\bibfnamefont
  {J.~C.}\ \bibnamefont {{S{\'e}amus Davis}}},\ }\bibfield  {title} {\enquote
  {\bibinfo {title} {{An Intrinsic Bond-Centered Electronic Glass with
  Unidirectional Domains in Underdoped Cuprates}},}\ }\href {\doibase
  10.1126/science.1138584} {\bibfield  {journal} {\bibinfo  {journal}
  {Science}\ }\textbf {\bibinfo {volume} {315}},\ \bibinfo {pages} {1380}
  (\bibinfo {year} {2007})},\ \Eprint
  {http://arxiv.org/abs/cond-mat/0703309}{cond-mat/0703309}\BibitemShut
  {NoStop}%
\bibitem [{\citenamefont {{Metlitski}}\ and\ \citenamefont
  {{Sachdev}}(2010)}]{MMSS10b}%
  \BibitemOpen
  \bibfield  {author} {\bibinfo {author} {\bibfnamefont {M.~A.}\ \bibnamefont
  {{Metlitski}}}\ and\ \bibinfo {author} {\bibfnamefont {S.}~\bibnamefont
  {{Sachdev}}},\ }\bibfield  {title} {\enquote {\bibinfo {title} {{Quantum
  phase transitions of metals in two spatial dimensions. II. Spin density wave
  order}},}\ }\href {\doibase 10.1103/PhysRevB.82.075128} {\bibfield  {journal}
  {\bibinfo  {journal} {Phys. Rev. B}\ }\textbf {\bibinfo {volume} {82}},\
  \bibinfo {eid} {075128} (\bibinfo {year} {2010})},\ \Eprint
  {http://arxiv.org/abs/1005.1288}{arXiv:1005.1288
  [cond-mat.str-el]}\BibitemShut {NoStop}%
\bibitem [{\citenamefont {{Sachdev}}\ and\ \citenamefont {{La
  Placa}}(2013)}]{SSRLP13}%
  \BibitemOpen
  \bibfield  {author} {\bibinfo {author} {\bibfnamefont {S.}~\bibnamefont
  {{Sachdev}}}\ and\ \bibinfo {author} {\bibfnamefont {R.}~\bibnamefont {{La
  Placa}}},\ }\bibfield  {title} {\enquote {\bibinfo {title} {{Bond Order in
  Two-Dimensional Metals with Antiferromagnetic Exchange Interactions}},}\
  }\href {\doibase 10.1103/PhysRevLett.111.027202} {\bibfield  {journal}
  {\bibinfo  {journal} {Phys. Rev. Lett.}\ }\textbf {\bibinfo {volume} {111}},\
  \bibinfo {eid} {027202} (\bibinfo {year} {2013})},\ \Eprint
  {http://arxiv.org/abs/1303.2114}{arXiv:1303.2114
  [cond-mat.supr-con]}\BibitemShut {NoStop}%
\bibitem [{\citenamefont {{Chowdhury}}\ and\ \citenamefont
  {{Sachdev}}(2014)}]{DCSS14b}%
  \BibitemOpen
  \bibfield  {author} {\bibinfo {author} {\bibfnamefont {D.}~\bibnamefont
  {{Chowdhury}}}\ and\ \bibinfo {author} {\bibfnamefont {S.}~\bibnamefont
  {{Sachdev}}},\ }\bibfield  {title} {\enquote {\bibinfo {title} {{Density-wave
  instabilities of fractionalized Fermi liquids}},}\ }\href {\doibase
  10.1103/PhysRevB.90.245136} {\bibfield  {journal} {\bibinfo  {journal} {Phys.
  Rev. B}\ }\textbf {\bibinfo {volume} {90}},\ \bibinfo {eid} {245136}
  (\bibinfo {year} {2014})},\ \Eprint
  {http://arxiv.org/abs/1409.5430}{arXiv:1409.5430
  [cond-mat.str-el]}\BibitemShut {NoStop}%
\bibitem [{\citenamefont {{Chowdhury}}\ and\ \citenamefont
  {{Sachdev}}(2015)}]{DCSS15b}%
  \BibitemOpen
  \bibfield  {author} {\bibinfo {author} {\bibfnamefont {D.}~\bibnamefont
  {{Chowdhury}}}\ and\ \bibinfo {author} {\bibfnamefont {S.}~\bibnamefont
  {{Sachdev}}},\ }\bibfield  {title} {\enquote {\bibinfo {title} {{Higgs
  criticality in a two-dimensional metal}},}\ }\href {\doibase
  10.1103/PhysRevB.91.115123} {\bibfield  {journal} {\bibinfo  {journal} {Phys.
  Rev. B}\ }\textbf {\bibinfo {volume} {91}},\ \bibinfo {eid} {115123}
  (\bibinfo {year} {2015})},\ \Eprint
  {http://arxiv.org/abs/1412.1086}{arXiv:1412.1086
  [cond-mat.str-el]}\BibitemShut {NoStop}%
\end{thebibliography}%

\end{document}